\documentclass[twocolumn]{pasj00}
\draft

\begin{document}
\SetRunningHead{Takagi, Vansevi\v cius, \& Arimoto}
{SEDs of dusty galaxies}

\Received{yyyy/mm/dd}
\Accepted{yyyy/mm/dd}

\title {Spectral Energy Distributions of Dusty Galaxies}

\author{
Toshinobu \textsc{Takagi}\altaffilmark{1,2,3}
Vladas \textsc{Vansevi\v cius}\altaffilmark{4}
Nobuo \textsc{Arimoto}\altaffilmark{5,6}}
\altaffiltext{1}{ 
Department of Physics, Rikkyo University,
 3-34-1 Nishi-Ikebukuro, Toshima-ku, Tokyo 171-8501
}
\altaffiltext{2}{ 
 The Institute of Space and Astronautical Science, 
 3-1-1 Yoshinodai, Sagamihara, Kanagawa 229-8510
}
\altaffiltext{3}{ 
 The Blackett Laboratory, Astrophysics Group,
 Imperial College, Prince Consort Road, London SW7 2BZ, UK
}
\email{t.takagi@ic.ac.uk}

\altaffiltext{4}{
 Institute of Physics, Go\v stauto 12, Vilnius 2600, Lithuania
} 
\altaffiltext{5}{ 
 National Astronomical Observatory of Japan
 2-21-1, Osawa, Mitaka, Tokyo 181-8588
}
\altaffiltext{6}{ 
 Institute of Astronomy, School of Science, University of Tokyo
 2-21-1, Osawa, Mitaka, Tokyo 181-0015
}


%


\maketitle

\begin{abstract}
We present a SED model of dusty galaxies, in which the equation of
radiative transfer is solved by assuming spherical symmetry.
The temperature fluctuation of very small dust particles is calculated
consistently with the radiative transfer.
The adopted dust model consists of graphite and silicate grains and PAHs,
whose relative fractions are
determined for each MW, LMC and SMC type extinction curve.
This model allows us to derive the intrinsic SEDs of stellar populations
embedded in dusty ISM, which are very important indicators for the age
of stellar populations. Therefore, the evolutionary phase of starburst
galaxies which have frequently very dusty ISM
can be investigated with this SED model.
We show that the SEDs of Arp220 and M82 can both be explained by the
same single stellar population, despite the significant differences
in the SEDs and the infrared luminosities.
The apparent difference between their SEDs
is mainly caused by the difference in the optical depth.
In contrast, the SED of prototypical star-forming ERO, HR10,
indicates that this galaxy is relatively old
comparing to Arp220 and M82.
It is found that, in the case of optically thin limit like
elliptical galaxies, the optical depth cannot be inferred
only from the SED, due to a degeneracy between the optical depth,
galactic size, and the spatial distribution of dust; the latter two are
important for estimating the average temperature of dust grains in 
elliptical galaxies.
When the observed size of elliptical galaxies is adopted for
the model geometry,
SEDs can be used to constrain the spatial
distribution of dust in elliptical galaxies.
\end{abstract}


\section{Introduction}
Age estimate of galaxies is one of the most important and essential tasks
in modern astronomy, although difficult.
Recent successive findings of high redshift galaxies
allow age estimates of young
galaxies to understand their formation processes. In this context,
spectrophotometric evolutionary models of galaxies have been widely
applied to derive
ages of the high redshift galaxies such as 53W002 (Windhorst et al. 1991),
B2 0902+34 ( Lilly 1988;  Bithell \& Rees 1990; Windhorst et al. 1991;
Eisenhardt \& Dickinson 1992), 4C41.17 (Chambers \& Charlot 1990; Chambers,
Miley \& van Breugel 1990; Windhorst et al. 1991; Graham et al. 1994;
Lacy \& Rawlings 1996; Dey et al. 1997), 6C1232+39 (Eales et al. 1993),
B2 0920+34 (Chambers \& Charlot 1990), and B2 1056+39 (Bithell \& Rees 1990).
However, the adopted photometric data for spectral energy distributions (SEDs)
are mainly restricted to the UV-optical
range in the rest frame, except for few galaxies.
It turned out that in such cases the SED fitting
is less helpful in order
to determine uniquely the ages of these galaxies.
This difficulty arises from the fact that the UV-optical SEDs of
young stellar populations heavily obscured by dust are very similar
to those of intrinsically old stellar populations; in other words,
the UV-optical SEDs are degenerate in age and dust attenuation.
In order to estimate the age of galaxies which contain a large
amount of dust, it is necessary to analyse SEDs within a wide range
of wavelengths including the dust emission
(Takagi, Arimoto \& Vansevi\v cius 1999).

The importance of detailed modelling of dust effects on SEDs
of galaxies has been demonstrated recently (e.g. Mazzei, De Zotti \& Xu 1994;
Gordon, Calzetti \& Witt 1997; Vansevi\v cius, Arimoto \& Kodaira 1997;
Silva et al. 1998; Efstathiou, Rowan-Robinson \& Siebenmorgen 2000).
However, the interpretation of the SEDs of dusty galaxies in the UV-NIR,
which are the key to estimate the age of the stellar population (e.g.
Takagi et al. 1999), is controversial.
In the SED model by Silva et al. (1998), the SEDs of Arp220 and M82
in the UV-NIR are totally dominated by the relatively old stellar
populations; i.e. unrelated to the very young stellar populations which
are newly formed in recent starburst events. On the other hand,
Efstathiou et al. (2000) show that the UV-NIR SED of M82 can be
explained by the obscured young stellar populations.
Although Silva et al. (1998) and Efstathiou et al. (2000) adopt
the different photometric data with the different aperture size, this
difference is not enough to explain the contradiction.
Silva et al. (1998)
divided the interstellar matter (ISM) of model galaxy into the
star forming
molecular clouds (MCs) and the diffuse medium.
The formation process of stars
is simulated, provided that stars are born only within the optically
thick MCs and
progressively escape from them.
When the elapsing time from the onset of starburst is
shorter than the time scale for stars to escape from MCs, all newly born
stars are embedded in MCs and their light is completely hidden at the optical
wavelengths, as they found for Arp220 and M82.
Therefore, in their model, the contribution from
starburst stellar populations to UV-NIR SEDs
is negligible for any aperture size of
these galaxies.
Note that Efstathiou et al. (2000) derive the age of starburst by applying
the SED model, while Silva et al. (1998) assume the age of starbursts
to reproduce observed SEDs.

The following observational results provide important clues to the origin
of the UV-NIR light of starbursts:
1) Rakos, Maindl \& Schombert (1996) showed that the UV-optical 
SEDs of starburst
galaxies are dominated by strongly reddened, young stellar populations;
2) Schmitt et al. (1997) demonstrated that UV-NIR SEDs of starburst
galaxies are different from those of normal galaxies;
3) Meurer et al. (1997) showed a tight correlation between the UV spectral
slope and the ratio of FIR flux to UV flux;
4) Goldader et al. (1995) showed that the 2 $\mu$m continua of the ten
ultraluminous infrared galaxies are dominated by red supergiants
in starburst stellar populations,
by using the CO indices of these galaxies.
All these observations indicate a significant contribution of the starburst
population to the UV-NIR SEDs of starburst galaxies.
Therefore, in this paper, the UV-NIR SEDs  
consistently coupled with
FIR-submm SEDs are used to estimate the age of starburst galaxies.
In this way, it should be possible to reduce the uncertainties of 
estimated age considerably. 

The SED of starburst galaxies has sometimes been modelled
by using the empirical relation between IRAS colours and total IR luminosities
(e.g. Devriendt, Guiderdoni \& Sadat 1999; Totani \& Takeuchi 2002).
We would like to point out a vital importance of
solving the radiative transfer problem in SED modelling, 
since it is the only method to employ the relations between 
the SED features in the whole wavelength range and physical 
parameters of the galaxies and dust consistently. 
Therefore, some additional spectral information is very helpful 
in order to constrain physical parameter space allowed for fitting, 
e.g. the silicate absorption feature at 9.7 $\mu$m arising due to 
the self-absorption by dust, can be used to constrain 
the type of extinction curve in our model. 

Here, we present a SED model of dusty galaxies by solving the equation of
radiative transfer,
in which stars are diffusely distributed within a spherical
region. This assumption for geometry is clearly different from
those by Efstathiou et al. (2000) in which
a starburst region is modelled as an ensemble of star-forming
molecular clouds.
The Hubble Space Telescope (HST) revealed that
most of nearby starburst galaxies actually have clusters of
newly formed massive stars
(O'Connell, Gallagher \&  Hunter 1994; Meurer et al. 1995;
Tacconi-Garman, Sternberg \& Eckart 1996),
so-called super star clusters (SSCs).
However, Meurer et al. (1995) reported that the SSCs contribute
only 20\% of the UV luminosity on average.
Thus, UV light from starburst region is dominated 
by the light from diffusely distributed
stars as modelled here, rather than that from star clusters.
Also, note that Shioya, Taniguchi \& Trentham (2001) suggest that 
in Arp220 the contribution from SSCs to the bolometric 
luminosity is at most 20\%. 

Even if the observed SED of a galaxy is
reproduced by the SED models,
it is important to verify the uniqueness of the solution.
Generally, a sophisticated handling of the SED model of
dusty galaxies requires a number of parameters. In such a case,
all the parameters may not be determined from the SED fitting alone;
thus, the solution may not be unique.
Therefore, it is important to investigate the dependence of
SEDs on each parameter. Once these test results are obtained,
it becomes clear which parameters can be determined from the SED fitting.

In this paper, we first intend to study the basic properties of 
SEDs of dusty galaxies, which are helpful to have an 
insight on what physical parameters can 
be constrained with the specific properties of SEDs. 
Then, we apply our SED model to dusty 
galaxies, a representative of the most active starburst (Arp220), 
a nearby less luminous starburst (M82), a distant extremely red 
star-forming galaxy (HR10), and an elliptical galaxy (NGC2768) as
a study for optically thin case. 
Specifically, we focus on the starburst age of dusty starbursts 
derived from UV-submm SEDs. 

This paper is organized as follows.
The model description and the SED analysis of parameter sensitivity
are given in \S 2 and \S 3, respectively. Then, we apply our model
to the observed SEDs in \S 4. 
Discussion and conclusions are given in \S 5
and \S 6, respectively. The cosmological parameters adopted throughout
this paper are $H_{0} = 75$ km s$^{-1}$ Mpc$^{-1}$,
$\Omega=1$ and $\Lambda=0$.

\section {Model}

We have developed a code which solves the equation of radiative
transfer by assuming spherical symmetry for arbitrary radial distributions
of stars and dust.
Isotropic multiple scattering is assumed and the self-absorption of
re-emitted energy from dust is fully taken into account.
The adopted dust model takes into account graphite and
silicate grains, and polycyclic aromatic hydrocarbons (PAHs).
We successfully reproduce the typical extinction curves in Milky Way (MW),
Large and Small Magellanic Cloud (LMC and SMC) as well
as the spectra of cirrus emission in the MW, which are modelled by
considering the temperature fluctuation of very small grains and PAHs.
As an input to our modelling, the intrinsic stellar SEDs of galaxies are taken
from Kodama \& Arimoto (1997).
We assume that the SEDs of galaxies are dominated by the starburst region or
by the diffusely distributed stars and dust: the former case is for starburst
galaxies and the latter one for elliptical galaxies.
In both cases the adopted geometry is defined by the radial distribution
of stars and dust with different scale lengths.

\subsection {Stellar Spectral Energy Distribution of Galaxies}

An evolutionary synthesis code developed by Kodama \& Arimoto (1997),
which gives consistent chemical and spectral evolution of a galaxy, 
is used for generating the intrinsic stellar SEDs. The basic structure of
the code follows Arimoto \& Yoshii (1986) prescription, i.e. the effects
of stellar metallicity are explicitly taken into account in calculating
galaxy spectra. The characteristics of this model are as follows. 
\begin{itemize}
\item New stellar evolutionary tracks are
incorporated comprehensively (VandenBerg et al. 1983;
Bressan, Chiosi, \& Fagotto 1994; Fagotto et al. 1994a,
1994b) and late stellar evolutionary stages
(horizontal branch - Yi 1996; asymptotic giant branch [AGB] -
Vassiliadis \& Wood 1993; post-AGB and white dwarf -
Vassiliadis \& Wood 1994) are fully included 
\item For the stellar spectra, Kurucz's (1992) stellar flux library is 
used for the metallicity $0.0001 \le Z \le 0.05$, the effective 
temperature $4000~ {\rm K} \le T_{\rm eff} \le 50000~ {\rm K}$, 
and the surface gravity $0.0 \le \log g \le 5.0$. 
The library covers the wavelength range from $9~n$m
to $3.2~ \mu$m with the spectral resolution $\Delta \lambda =1~n$m - 
$2~n$m 
for $\lambda \le 1~ \mu$m and $\Delta \lambda= 5~n$m - $10~n$m for
$\lambda > 1~ \mu$m. 
\item  The spectra of cool stars with 
$T_{\rm eff} \le 4000~ {\rm K}$ are taken from
Pickles (1985) and Gunn \& Stryker (1983). 
\end{itemize}

Throughout this paper, we adopt the model of simple stellar
populations (SSPs) with solar metallicity 
as the intrinsic stellar SED model, otherwise noted explicitly.  
We adopt the Salpeter initial mass function (IMF) with
stellar masses ranging from $0.1~M_\odot$ to $60~M_\odot$. 
Some examples of SEDs of SSPs are shown in Figure 1.
In general, the intrinsic SEDs of
stellar populations are mainly determined
by both the age and the star formation history.
Therefore, in order to derive the definite age of
stellar populations from the SED, it is necessary to determine
the time scale of star formation, which can be expressed in terms
of the dynamical time, sound-crossing time, and cooling time.
In this study, we mainly
focus on starburst galaxies, in which the variation
of stellar populations is less significant, comparing to
normal spiral galaxies. Even in such an extreme case,
it is difficult to know the characteristic time
scale of each starburst event. 
Therefore, we choose the simplest stellar population 
as an indicator of the age of starbursts, rather than 
try to determine precise ages of starbursts from intrinsic stellar SEDs. 

Gas emission is not taken into account in our model which is designed to
calculate the continuum emission of galaxies.
A modification of SEDs
due to the gas emission is not significant, unless a galaxy is younger
than 10 Myr for SSPs
(Leitherer \& Heckman 1995; Fioc \& Rocca-Volmerange 1997).
As we will show later, our model can reproduce the observed SEDs 
of dusty galaxies with the age of SSP older than 10 Myr. 
Therefore, we believe that the effect of gas 
emission does not affect our conclusions.

\subsection{Dust model}

\subsubsection{Optical Properties}

The optical properties of graphite and silicate are taken from Laor \&
Draine (1993).
We adopt a "smoothed astronomical silicate" which is obtained by removing the
absorption feature at $1/\lambda=6.5~ \mu$m$^{-1}$ 
from the imaginary part of the
original dielectric function
(Draine \& Lee 1984; Laor \& Draine 1993; Weingartner \& Draine 2001).

Optical properties of PAH are adopted according to L\'eger,
d'Hendecourt \& D\' efourneau (1989), and D\'esert, Boulanger \& Puget (1990).
The PAH features are assumed to
have the Lorentz profile, which successfully
reproduced the spectra of the reflection nebula NGC7023 and the Ophiuchus
molecular cloud observed by ISO (Boulanger et al. 1998).
The Lorentz function is given by:

\begin{equation}
F(\nu)=Af(\nu) = \frac{A}{\pi \Gamma}
\left [1+ \left (\frac{\nu - \nu_0}{\Gamma} \right )^2 \right ]^{-1},
\end{equation}
where $f(\nu)$ is the normalized Lorentz profile, 
and $A$ is the integrated cross-section, and 
$\nu_0$ and $\Gamma$ give the frequency and the 
width of the band, respectively.
The adopted parameters
for the PAH features at $\lambda$ = 3.3, 6.2, 7.7, 8.6, 11.3 and $12.7~ \mu$m
are given in Table 1.
The number of H atoms attached to a PAH molecule is given by
$N_{H}= x_H (6N_C)^{1/2}$, where $x_H$ ($\le 1$ by definition)
is the degree of dehydrogenation, and
$N_C$ is the number of carbon atoms in one PAH molecule (Omont 1986).
Since the binding energy of C-H bond is weaker than that of
C-C bond, the H atoms tend to be removed in the strong radiation field.
This suggests that $x_H$ could vary with the strength of radiation field.
However, recent observations showed that relative intensities of PAH features are
independent of the strength of radiation field (Lemke et al. 1998; Chan et al. 2001).
Thus, it is too early to introduce
the relation between $x_H$ and the strength of radiation field,
instead we assume $x_H=0.4$ according to D\'esert et al. (1990).

\subsubsection{Size Distribution}
The spectra of diffuse high latitude clouds
(so-called Galactic cirrus), obtained by the IRAS all-sky survey (Low et al. 1984), show
excesses of emission at 12 and $25~ \mu$m over those expected from dust grains radiating at
the equilibrium temperature. Draine \& Anderson (1985) suggested that the excessive emission
at MIR could be radiated by grains small enough to undergo the temperature fluctuation in
the local interstellar radiation field, due to their small heat capacity.
Such very small grains (VSGs), typically smaller than $0.01~ \mu$m, are
transiently heated up to several hundreds of Kelvin by the interstellar
radiation field and emit in the MIR.
Hereafter, we define both graphite and silicate grains with the radii $a < 0.01~ \mu$m
and $a > 0.01~ \mu$m as VSGs and big grains (BGs), respectively.

For BGs, we adopt a power law size distribution derived by
Mathis, Rumpl \& Nordsiek (1977, hereafter MRN).
For VSGs, the same size distribution is adopted, but with the different index
(Draine \& Anderson 1985).
Introducing the radius $a_b = 0.01~ \mu$m, at which a break between the two power laws
takes place, the size distribution is given by:
\begin{equation}
dn_k =
\left\{ \begin{array}{ll}
n_H C_k a_b^{\beta} (a/a_b)^{\gamma_k} da,
& \quad \mbox{for $a_{min} < a < a_b$ (VSGs)} \\
n_H C_k a^\beta da,& \quad \mbox{for $a_b < a < a_{max}$ (BGs)}
\end{array}\right.
\end{equation}
where $dn_k$ is the number density of grains of type $k$ ($k=g$ for graphite and $k=s$ for
silicate) in the interval [$a,a+da$], $n_H$ is the number
density of H nuclei, and $C_k$ is the coefficient which determines the ratio of dust to
hydrogen.
For both graphite and silicate grains, we adopt $\beta =-3.5$,
$a_{min}=0.001~ \mu$m, $a_b = 0.010~ \mu$m and $a_{max}=0.25~ \mu$m.
The indices of VSG size distribution $\gamma_k$ are determined by
the fits to the MW extinction curve and the spectrum of Galactic cirrus.

For PAHs, we write the size distribution as a power law of $N_C$:
\begin{equation}
dn_{P} = n_H C_{P} N_C^{\gamma_{P}} dN_C,
\quad \mbox{for $N_{C1} < N_C < N_{C2}$},
\end{equation}
where $N_{C1}$ and $N_{C2}$ are the number of C atoms in the smallest and
largest PAH, respectively. We use the subscription $k=P$ for PAHs.
Since the small PAHs with $N_C < 20$ will evaporate even in the weak
radiation field like the local interstellar radiation field (Omont 1986),
$N_{C1}=20$ is adopted in this work.
On the other hand, $N_{C2}$ is less constrained.
$N_{C2}$ should be less than $\sim 1000$ so that
a single photon could heat up PAHs to the temperature at which the emission peak is located
at MIR (Puget \& L\' eger 1989). We assume that the smallest graphite grain and the largest
PAH contain the same number of C atoms (see also Dwek et al. 1997).
According to the relation between $a$ and $N_C$:
\begin{equation}
a (\mu \mathrm{m}) = \left \{ \begin{array}{ll}
 1.29 \times 10^{-4} N_C^{1/3}, & \quad \mbox{for graphite grains} \\
 10.0 \times 10^{-4} (N_C/120)^{1/2},
& \quad \mbox{for PAHs} \end{array}\right.
\end{equation}
we get $N_{C2}=465$ for $a_{min}=0.001~ \mu$m.
It is assumed that $\gamma_P$ is equal to the index of size distribution of graphite
VSGs, which is written as a function of $N_C$.
Using above relations between $a$ and $N_C$, $\gamma_{P}$ is given by
$(\gamma_g-2)/3$.

\subsubsection{Temperature Fluctuation}

When the dust grains attain temperature equilibrium with a
surrounding radiation field, $J_\lambda$, by radiating the same amount of energy
as the absorbed one, the equilibrium temperature, $T_{eq}$, is determined by:
\begin{equation}
4 \pi \int \sigma^{a}_{\lambda ,k} (a) J_\lambda d\lambda = 4 \pi \int
\sigma^{a}_{\lambda ,k} (a) B_\lambda(T_{eq}) d\lambda,
\end{equation}
where $\sigma^{a}_{\lambda ,k} (a)$ is the absorption cross-section of dust grains of
type $k$ and size $a$, and $B_\lambda$ is the Planck function.

When dust particles are small enough, grains cannot attain the temperature
equilibrium with the interstellar radiation field.
In this case, we have to calculate the temperature distribution function
$dP/dT$, which is the probability for finding a particle in the interval $[T,T+dT]$,
in order to predict the emission spectra.
Methods to calculate $dP/dT$ have been presented by several
authors (Draine \& Anderson 1985; D\'esert, Boulanger \& Shore 1986; Dwek 1986;
Guhathakurta \& Draine 1989). We follow the method outlined by Guhathakurta \& Draine (1989),
in which the transition matrix is defined as the probability per unit time of a particle
making a transition in the enthalpy space by the emission or absorption, and the solution
is derived by inverting this transition matrix. The heat capacities of graphite and silicate
are also taken from Guhathakurta \& Draine (1989). For PAHs, the same heat capacity is used
as that of graphite. Although the different expressions are given for its heat capacity
(e.g. L\'eger et al. 1989; Dwek et al. 1997),
alternative choice has no significant influence
on the emission spectra (Siebenmorgen \& Kr\"ugel 1992).
By using $dP/dT$, the
emissivity per dust particle with the size $a$ and the constituent $k$ is given by:
\begin{equation}
\varepsilon_{\lambda ,k} (a) = 4 \pi \sigma_{\lambda ,k}^a (a) \int_0^{T_{sub,k}}
B_\lambda (T) \frac{dP_k}{dT}(a,T) dT,
\end{equation}
where the sublimation temperature $T_{sub,k}$ = 800 K 
is assumed for all the dust
constituents 
according to Carico et al. (1988) who showed that the emission from
hot-dust particles with the temperature of $\sim 800$ K contributes to
the NIR luminosity of IRAS bright galaxies. This assumption is safe
because, unless $T_{sub,k}$ is higher than 800 K and the starburst is
younger than 10 Myr, the resulting SEDs depend little on $T_{sub,k}$.
If we adopt $T_{sub,k}$=1500 K (c.f. Kobayashi et al. 1993), 
a little excess of emission ($<10 \%$) can be seen in the NIR 
wavelengths only for the optically thick cases in which dust emission 
dominates the NIR emission, and therefore this effect is negligible 
to the results of SED fitting. 
In case of the temperature equilibrium, $dP/dT$ is given by the delta
function $\delta (T-T_{eq})$.

\subsubsection{Dust in the MW, LMC and SMC}

Figures 2 and 3 show the model fits to the extinction curve and the spectrum of cirrus
emission in the MW, respectively.
The extinction curve is given by the cross-section per hydrogen atom:
\begin{equation}
\sigma^H_\lambda = \frac{1}{n_H} \sum_k \int \sigma^e_{\lambda ,k}(a) \frac{dn_k}{da} da,
\end{equation}
where $\sigma^e_{\lambda ,k}(a)$ is the extinction-cross section of the type $k$
and size $a$ dust grains.
The adopted interstellar radiation field is taken from Mathis, Mezger \& Panagia (1983).
The contribution of PAHs to the spectrum and the extinction curve is significant;
they dominate the spectrum at $\lambda < 12~ \mu$m,
and are responsible for the nonlinear extinction
rise at FUV. The predicted extinction at FUV is a factor of two higher than the average
extinction curve in the MW. The dust model presented by
Dwek et al. (1997), in which the same optical properties are adopted, also overestimates
the FUV extinction curve.
Unfortunately, as far as the visible-UV absorption of PAHs in
the gas phase is concerned, very few studies are available especially for the large PAHs
(Verstraete \& L\'eger 1992). We think, this discrepancy should be solved by using the
more reliable optical properties of PAHs than those available at present.

Since the cirrus spectra of the LMC and SMC are not available at present, only the extinction
curves are used to derive relative grain fractions, provided that fraction
ratios of carbonaceous particles (graphite BGs, VSGs and PAHs) are the same as those of the MW
(see also Pei 1992). The model fits to the extinction curve in the LMC and SMC are also
shown in Figure 2. For normalisation of the extinction curves, the adopted ratios of
hydrogen column density to colour excess $E(B-V)$ are 5.8 $10^{21}$, 2.0 $10^{22}$,
and 4.6 $10^{22}$ for the MW, LMC, and SMC, respectively (Mathis 1990). In Table 2,
we give the derived mass
ratios of dust to hydrogen and the relative mass fractions of various constituents of dust.
Note that the fraction of VSGs of type $k$, $f_k^{VSG}$ and that of BGs,
$f_k^{BG}$, are related by:
\begin{equation}
\frac{ f_k^{VSG}}{f_k^{BG}} = \frac{\beta+4}{\gamma_k+4}
\frac{a_b^{\gamma_k+4}-a_{min}^{\gamma_k+4}}
{a_{max}^{\beta+4}-a_{b}^{\beta+4} }.
\end{equation}
It is difficult to determine the amount of silicate VSGs, because they have
little influence on both of the extinction curve and the spectrum of cirrus emission.
Therefore, we simply extend downward the lower end of MRN distribution from $a_b$ to
$a_{min}$ for silicate, i.e. $\gamma_s =\beta$.
For graphite VSGs, we derive $\gamma_g = -3.75$.

\subsection {Radiative Transfer}

\subsubsection {Geometry}
The density distributions, $\rho(r)$, of stars and dust are
assumed to be given by King's law:
\begin{equation}
\rho(r)=\frac{\rho_0}{\left( 1+\left( \frac{\displaystyle r}
{\displaystyle %
r_{c}}\right) ^{2}\right) ^{\frac{3}{2}}} ,
\end{equation}
where $\rho_{0}$ is the density at the centre of galaxy and $r_{c}$
is a core radius. We fix a geometry of stars and
change the dust distribution by choosing a value of parameter
$\eta \equiv r_{c,D}/r_{c,S}$, where $r_{c,S}$ and $r_{c,D}$ are
the core radius of stars and dust, respectively. 

The second parameter that defines the geometry is a tidal radius $r_t$ which
determines $\rho_0$ with a given total mass.
When the total mass of stars is fixed, the temperature of dust is controlled
by the tidal radius of stars $r_{t,S}$, 
since the energy density of the radiation field is approximately
proportional to the stellar density. In order to keep a source function
(see below) constant, we adopt a mass-radius relation:
\begin{equation}
r_{t,S}= 10 R \left ( \frac{M_*}{10^{11} M_\odot } \right )^{0.5}
\quad \mbox{[kpc]},
\end{equation}
where $M_*$ is the total stellar mass and $R$ is a scaling factor, and
the normalisation is given for the scale of typical galaxy.
Note that, for constant value of $R$,
the relative shape of SED is conserved for different
$M_*$ due to this relation, when the optical depth is kept constant.
We assume that the tidal radius of dust $r_{t,D}$ is the same as 
$r_{t,S}$, i.e. $r_{t,D}$=$r_{t,S}$(=$r_t$ hereafter), and there is 
no dust at $r > r_t$. 
We adopt the concentration parameter $c\equiv \log (r_{t}/r_{c,S})=2.2$,
which is the typical value for normal elliptical galaxies
(Combes et al. 1995). 
This value is also adopted for 
starburst galaxies in which the real distribution of stars 
inside the dominant starburst regions is difficult to derive 
due to the effect of dust attenuation. However, the effects 
of $c$ on the SED are marginal, and detectable only at MIR region, 
and therefore have negligible impact on the values of 
parameters derived in \S 4.

The dust distributions
$\rho_{D}(r)$ for $\eta=1$, 10, 100, and 1000 are shown in Figure 4. 
Note that, for larger values of $\eta$, $r_{t}$ can be smaller than 
$r_{c,D}$, due to the definition given above. 
In the case of $\eta =1000$, dust distributes nearly uniformly
and surrounds centrally concentrated stars,
while in the case of $\eta =1$, stars and dust are distributed identically.
The former case is similar to the shell geometry
suggested for starburst galaxies by Gordon, Calzetti \& Witt (1997).
However, note that the shell geometry gives SEDs different
from our case if $\eta =1000$ and $\tau_V \gg 1$, since the shell geometry
results in the spectral cut-off around NIR
(see the results by Rowan-Robinson \& Efstathiou 1993).

A mass of dust, $M_D$, can be derived as follows.
The optical depth at V band is defined by:
\begin{equation}
\tau_V = \int_0^{r_t} n_H(r) \sigma^H_V dr ,
\end{equation}
where $n_H(r)$ is the number density of hydrogen and $\sigma^H_V$ is
the extinction cross-section of dust per hydrogen at V band.
A total number (or total mass) of hydrogen is derived by
using the distribution function defined by $\eta$,
once $\tau_V$ is specified.
The mass ratio of dust to hydrogen is determined by the dust model.
Since the column density is proportional to $r_t^{-2}$ for
a given mass, $M_D$ is given by:
\begin{equation}
M_D = M_{0} \frac{\tau_V}{\tau_0} \left ( \frac{r_t}{10 \mbox{~kpc}}
\right )^2,
\end{equation}
where both $M_{0}$ and $\tau_0$ are the normalisation factor.
By using the dust model described in \S 2.2 for the MW, LMC and SMC,
$M_{0}$ is found to be $5.5 \ 10^7 M_\odot$,
$9.4\ 10^7 M_\odot$ and $1.1\ 10^8 M_\odot$, respectively.
The second normalisation factor $\tau_0$ depends on the geometry and is
equal to 1.0, 1.78, 33.86, and 1753 for
$\eta = 1000$, 100, 10, and 1, respectively.

\subsubsection {Method of Calculation}
Solving the equation of radiative transfer for an arbitrary radial
distribution of stars and dust, we adopt the "method of rays" (Band \& Grindlay 1985).
The schematic picture of this method is shown in Figure 5.

It is convenient to divide the specific intensity, $I_\lambda$, into
three components,
\begin{equation}
I_\lambda = I^{(1)}_\lambda+ I^{(2)}_\lambda + I^{(3)}_\lambda,
\end{equation}
where the superscripts (1), (2) and (3) represent direct light from
stars, scattered light by dust grains, and thermal emission
from dust grains, respectively (Rowan-Robinson 1980).
In the case that the optical properties
of dust grains are independent of their temperature, each component of
intensity can be calculated independently in the order of the
superscript.
The details of iterative calculation of $I^{(2)}_\lambda$ and
$I^{(3)}_\lambda$ are discussed in the following subsections.

The differential equation for the specific intensity, $I_{\lambda }$,
along a ray is given by:
\begin{equation}
\frac{dI_{\lambda }(b,s)}{ds}=-n_H(r)\sigma _{\lambda }^{H}[I_{\lambda
}(b,s)-S_{\lambda }(r)].
\label{eq:rt}
\end{equation}
The impact parameter to the centre, $b$, and path length along
the ray, $s$, are related to the radial distance, $r$, according to an
expression:
\begin{equation}
r^{2}=b^{2}+(\sqrt{r_{t}^{2}-b^{2}}-s)^{2}.
\end{equation}
The source function for $I_{\lambda }^{(1)}$ is given by:
\begin{equation}
S_{\lambda }^{(1)}(r)=\frac{\varepsilon ^*_{\lambda }(r)}{4\pi
n_H(r)\sigma_{\lambda }^{H}},
\end{equation}
where $\varepsilon ^*_{\lambda }(r)$ is the emissivity of stellar
component at wavelength $\lambda$.

Spherical geometry of the model requires only radial grid to be specified.
The grid is defined by:
\begin{equation}
r_j = \left ( \frac{j-1}{N-1} \right )^d r_t,
\end{equation}
where $N$ is the total number of radial shells and $d$ is
a parameter that depends on the dust distribution.
The required $N$ depends on the adopted $\tau_V$. In all analysed cases,
the grid with $d=3$ and $N=160$ gave results with satisfying accuracy;
i.e. the difference between the input and output energy is less than
5\% (see \S 2.3.4).
If the required accuracy is violated the finer grid should be redefined.

For each $b_j$ (= $r_j$),
$I^{(1)}_\lambda$ is calculated at the crossing points of the ray
and shells' boundaries.
The grid of path length, $s_i$, is defined at these points.
Then, the equation for $j$-th ray is written as:
\begin{eqnarray}
I^{(1)}_\lambda (b_j, s_{i+1}) = \bar{S}_{i} + e ^{-\Delta \tau_i }
(I^{(1)}_\lambda (b_j, s_{i}) - \bar{S}_{i} ),  \label{eq:in}
\end{eqnarray}
with the boundary condition $I^{(1)}_\lambda (b_j, s_1) = 0$,
where $\Delta \tau_i = (s_{i+1}-s_i) \bar{n}_H \sigma^H_\lambda$, and
the bar above variables indicates an average of $i$-th and $(i+1)$-th values.
Once $I_\lambda (b_j, s_i)$ is calculated for all $j$ ($1<j<N$) and $i$
($1<i<2(N-j)+1$), one
can derive the mean intensity in case of spherical geometry:
\begin{eqnarray}
J^{(1)}_\lambda (r) = \frac{1}{2} \int_{-1}^{+1} I^{(1)}_\lambda (r, \mu )
d\mu,
\end{eqnarray}
or
\begin{eqnarray}
J^{(1)}_\lambda (r_{j'}) = \frac{1}{2} \sum_{j=1}^{j'-1}
(\bar{I}^{(1)}_\lambda(b_{j},
s_{j+}) + \bar{I}^{(1)}_\lambda(b_{j}, s_{j-})) \Delta \mu_{j},
\end{eqnarray}
where $\Delta \mu_{j}$ is the step of cosine grid, $\mu_{j}$,
which is defined by the angle between the vector pointing the direction of
the $j$-th ray through the radial point $r_{j'}$
({\boldmath $v$} in Figure 5) and the radius vector;
$s_{j+}$ and $s_{j-}$ are the path lengths of the
ray defined by $b_{j}$, from both sides of the outer edge to
the shell $r_{j'}$, which means $j+=N+(j'-1)-2(j-1)$ and $j-=N-(j'-1)$;
the bar above $I_\lambda$ means average of the $j$-th and $(j+1)$-th
values.

The components, $I^{(2)}_\lambda$ and $I^{(3)}_\lambda$, are
calculated in a similar manner by using the source function
characteristic for each component.

\subsubsection {Scattered Light}

The source function of $n$-times scattered light is:
\begin{eqnarray}
S_{\lambda ,n}^{(2)}(r,\mbox{\boldmath $n$})=\frac{\omega _{\lambda
}}{4\pi }%
\int g(\mbox{\boldmath $m,n$})I_{\lambda ,n-1}^{(2)}(r,\mbox{
\boldmath $m$}%
)d\Omega,
\end{eqnarray}
where $I_{\lambda ,n-1}^{(2)}$ is the intensity of $(n-1)$-times
scattered light,
$g(\mbox{\boldmath $m,n$})$ is the angular phase function for coherent
scattering from direction {\boldmath $m$} to {\boldmath $n$},
and $\omega _{\lambda }$ is a dust albedo (e.g. Mihalas 1978).
Isotropic scattering is assumed for simplicity,
that means $g(\mbox{\boldmath $m,n$})\equiv 1$.
The total intensity of scattered light, $I_{\lambda }^{(2)}$ is
given by $\sum_{n}{I_{\lambda ,n}^{(2)}}$. Since $I_{\lambda }^{(1)}$ can
be regarded as the 0-times scattered light, the source function of
once scattered light in the case of spherical geometry is:
\begin{equation}
S_{\lambda ,1}^{(2)}(r)=\frac{\omega _{\lambda }}{4\pi }\int
I_{\lambda ,0}^{(2)}(r,\mbox{\boldmath $n$})d\Omega =
\frac{\omega_{\lambda }}{2}\int_{-1}^{+1}I_{\lambda }^{(1)}(r,\mu )d\mu .
\end{equation}
Using this definition of the source functions,
the intensities of the higher terms of scattered light
are calculated by using the method given in the previous section.
When the number of scattering terms considered is not sufficient, this
results in a loss of the output energy.
We find that the losses are less than 0.3\% for the MW extinction
curve when up to six terms are taken into account.

In Figures 6 and 7, the mean intensity ratios of $n$ to $n-1$ times
scattered light as a function of $r$ are shown for different $\eta$ and
$\tau_V$ in the case of $\omega = 0.6$. Generally, the mean intensity
ratio approaches to a flat distribution for the higher
order of scattered light. For large $\tau_V$, this mean intensity ratio 
is hard to become totally constant from $r=0$ to $r_t$, due to the slow 
diffusion of photons. 
Figure 8 shows the fraction of the unabsorbed stellar light (direct light)
and the emerging scattered light both normalized with respect to the
total input luminosity for the uniform distribution of stars and dust
(Table 3A in Witt, Thronson \& Capuano 1992).
In our calculation, both values of the direct and scattered light
are systematically smaller in comparison with the results
obtained by Witt et al. (1992) who solved the radiative transfer
problem by using the Monte Carlo method.
The analytic solution for the direct light in an optically
thick case is given by Band \& Grindlay (1985) and is
plotted together.
Our results are consistent with the analytic solution. In Figure 9,
luminosity ratios of scattered to direct light are compared. The
results of Witt et al. (1992) are plotted by circles,
while ours are represented by solid lines.
Each solid line corresponds to a particular number of
scattering terms used for calculation of the scattered light.
Over the whole range of $\tau_V$, the results of Witt et al. (1992)
are consistent with ours when the first
five scattering terms are taken into account.

\subsubsection {Self-absorption}

In order to take into account a self-absorption of dust emission,
the radial distribution of $dP/dT$, is iteratively calculated for
every constituent and every discrete grain size until the energy
conservation is fulfilled.
The first guess of $dP/dT$, which is denoted as $dP/dT_1$,
is done using the mean intensity calculated from only
two components, $I_{\lambda }^{(1)}$ and $I_{\lambda }^{(2)}$.
With the assumption that the scattering of dust emission is negligible,
the source function for the $n$-th guess of the thermal component of intensity,
$I_{\lambda ,n}^{(3)}$, is given by:
\begin{equation}
S_{\lambda ,n}^{(3)}(r)= \frac{1}{n_H(r) \sigma^H_\lambda}
\sum_{k} \int \int_0^{Tsub,k} \sigma^a_{\lambda ,k} (a)
B_{\lambda }(T) \frac{dP_k}{dT_n} (a,T,r) dT \frac{dn_k}{da}da.
\label{eq:s3}
\end{equation}
Calculation of the $I_{\lambda ,n}^{(3)}$ is performed
in the same way as
described in \S 2.3.2. The sum of intensities, $I_{\lambda
}^{(1)}+I_{\lambda }^{(2)}+I_{\lambda ,n}^{(3)}$ is used to derive
 $dP/dT_{n+1}$. The mean temperature of dust rises step by step during the
iterative procedure. We continue the iteration until the energy conservation
is fulfilled.
Even in the worst cases ($\tau_V \sim 40$), the grid can be adjusted to keep
the "numerical" losses of the total energy below 5\%.
This numerical loss is less than 3\% in the nominal case.

The changes of the spatial distribution of equilibrium temperature of BGs,
$T_{eq}(r)$, and the SED due to
the self-absorption are shown in Figure 10, panels a and b, respectively.
Heating dust by the thermal emission of dust itself,
i.e. the self-absorption, has considerable effect on the
MIR emission if $\tau_V$ is larger than $\sim 10$.
The effect of self-absorption is most prominent in a case of
the SMC extinction curve since it has the largest absorption
cross section at MIR due to high fraction of silicate grains.

Ivezi\' c et al. (1997) have defined a set of benchmark
problems and gave exact solutions suitable for verification
of the radiative transfer codes dealing with dusty environments.
The benchmark problems are solved with a
central point source embedded in a spherically symmetric dust envelope
with an inner cavity being free of dust.
The density variation with $r$ is assumed to
be a power law, $\rho (r) \propto r^{-p}$.
We show the results of comparison with $p=0$
in Figure 11. The consistency is very good although, in the extreme
case of $\tau_{1\mu \mathrm{m}}=1000$, the temperature of dust is
slightly higher than the benchmark solution. By the definition of the
benchmark problem, the radius of inner cavity, $r_1$, has to be adjusted
during the iteration to keep the dust temperature at the inner edge of
dust envelope equal to 800 K. However,
we do not adjust the inner radius since we do not need such rigorous
adjustment for the purpose of galaxy modelling.
As a result, our temperature is slightly
higher ($\sim 830$ K) due to the heating by the
scattered light and dust emission.
This mutual heating of dust is
negligible when $\tau_{1 \mu \mathrm{m}} \le 100$.
On the other hand, in the case of $p=2$,
the effect of mutual heating is larger due
to more pronounced concentration of the dust
near the inner radius of the shell than the case of $p=0$.
In this case, the adjustment of the inner radius of dusty zone is
indispensable to compare the results properly with the benchmark solutions.

\section {Sensitivity Analysis}
It is necessary to know how the resulting SEDs depend on the model
parameters such as age $t$, optical depth $\tau_V$, core radius ratio
$\eta$, size scaling factor $R$, and type of the extinction curve,
in order to judge whether the results of SED fitting are unique or not.
In this section, we describe how the radial temperature distributions
and the SEDs depend on these parameters.
The standard model is defined by a set of parameters $t=10$ Myr,
$\tau_V=1.0$, $\eta=1000$, $R=1.0$, and the MW type extinction curve,
as tabulated in Table 3.
Unless otherwise stated, the standard model parameters are used
throughout this section.

\subsection {The Effect of Age}
Figure 12a shows the radial temperature distribution $T_{eq}(r)$ of
models for various ages of the system.
Due to high concentration of stars, the energy density of the radiation
field is highest at the centre, and the dust temperature attains maximum there.
As stellar populations aging, $T_{eq}(r)$ becomes systematically
lower in a whole system. As will be mentioned below,
this leads to a shift of dust emission toward longer wavelength.
At the same time, the temperature gradient becomes flatter,
since the effective wavelength of dust heating photons becomes longer.
Such photons can travel easier to the outer regions and heat the dust there.
However, this flattening of temperature distribution does not affect the SED,
significantly.

Figure 12b shows a behaviour of the SED as a function of age.
The SED of dust-rich objects is composed of
the stellar and dust emissions with the peaks in the UV-NIR
and MIR-submm ranges, respectively.
Both the stellar and dust emissions decrease as a system aging,
because both the emitted UV photons
and the absorbed ones by dust become less numerous.
When a system is young, the dust emission dominates the SED, while it
becomes less important for older ages.
Simultaneously, the dust becomes cooler and the dust emission peak
shifts from $\sim 40 \mu$m (for 10 Myr) to $100 \mu$m (for 10 Gyr).
Note that neither the dust absorption nor the dust emission affects
the NIR part of SED significantly,
while the fraction of energy emitted in FIR decreases monotonically as
a system aging. This indicates that the FIR luminosity normalized by
the H-band luminosity, $L_{FIR}/L_H$, is sensitive to age
(see also Takagi et al. 1999), where
$L_{FIR}$ is the FIR luminosity derived from 60 and 100 $\mu$m fluxes.

\subsection {The Effect of Optical Depth}

Figure 13a shows $T_{eq}(r)$ for different optical depths, $\tau_V$.
$T_{eq}(0)$ is almost independent of the optical depth for $\tau_V \le 40$.  
When a mean free path of the UV photon is larger than the core radius 
of stellar distribution, almost all stars in the system can 
contribute to the heating of dust at the system centre. 
Note that 
for the homogeneous distribution of dust the optical depth within 
$r_{c,S}$ is $10^{2.2}$ times smaller than that defined with $r_t$ in 
the adopted geometry. 
The SED of dust emission becomes broader with increasing $\tau_V$,
because the temperature at the outer region drops sharply, while 
$T_{eq}(0)$ is kept constant. This is not true for the case of 
$\eta \le 10$. In this case, $T_{eq}(r)$ is less sensitive to $\tau_V$, 
since strongly concentrated dust obscures the emission from the centre, and
the energy density in the outer region is determined by stars
in the outer region. 

The SEDs with various $\tau_V$ are shown in Figure 13b.
When $\tau_V$ increases, the stellar emission
decreases due to the dust absorption,
as a result of which the dust emission increases;
thus shows a remarkable contrast to the age effect.
The fraction of energy emitted in the FIR increases rapidly
as a function of $\tau_V$ when $\tau_V \ll 1$, 
but this fraction is less sensitive for $\tau_V \ge 1$.
Note that the peak wavelength of dust emission is less sensitive to
$\tau_V$ than to the age, since only the dust temperature
in the outer region depends on $\tau_V$,
while the age affects the dust temperature in a whole system.

It is worth noting that the SED approaches the source function
for large $\tau_V$ ($\gtrsim 10$).
In particular, this becomes conspicuous in the UV
region, since the extinction cross-section increases with decreasing
$\lambda$.
Thus, the SED in this region resembles the source function for large
$\tau_V$.
If stars distribute point-likely and are enshrouded by dust shell,
the source function, $S^{(1)}_\lambda$, in dusty medium is zero.
Therefore, there should be a sharp cut-off in the SED, i.e. the
emission in the UV-NIR spectral region of the population enshrouded
by dust is virtually negligible (e.g. Figure 5a in Rowan-Robinson 1980).

\subsection {The Effect of Dust Geometry}

Figure 14a shows the effect of dust geometry
$\eta=r_{c,D}/r_{c,S}$ on $T_{eq}(r)$.
$T_{eq}(0)$ drops slightly in the case of $\eta =1$,
since the mean free path of UV photons becomes comparable to $r_{c,S}$.
On the other hand, $T_{eq}(r)$ rises as $\eta$ decreases in the outer region,
since the optically thin region from which the UV photons 
can escape from a system enlarges.
However, this effect is rather minor when compared with the 
effect of age and $\tau_V$.

Figure 14b shows the effect of dust geometry on the SED.
The peak of dust emission moves toward shorter wavelength
for smaller $\eta$, although $T_{eq}(r)$ is not very sensitive to $\eta$.
This is because the fraction of dust located near the centre 
where the dust is hottest increases with decreasing $\eta$.
Note that even if $\tau_V$ is the same, the amount of dust changes when
$\eta$ is different.
When the dust mass with $\eta =1000$ is normalized to 1.0,
the value of dust mass becomes 0.6, 0.03 and 0.006 for $\eta =100,
\ 10$ and 1, respectively.
Since there is less amount of dust in a system with smaller $\eta$,
$L_{FIR}$ drops with decreasing $\eta$.
If the mass of dust is kept constant for different $\eta$, then
the $L_{FIR}$ is of the same order of magnitude.
In the cases of $\eta \ll 100$, the resulting SEDs are close to
the intrinsic SED. The bump feature at $2175$ \AA\ virtually disappears
because the emission from the optically thin region
dominates the resulting SED. 
Even if $\tau_V \gg 1$, the UV-NIR SED 
does not change significantly.
Thus, the observed spread of starburst galaxies in two-colour diagrams
can be explained by the stellar and dust distributions only with $\eta 
\gtrsim 100$ (cf. Gordon et al. 1997).
For $\eta \gtrsim 100$,
it is difficult to determine the value of $\eta$ precisely from the
SED fitting alone, since the resulting SEDs are degenerate in
geometry and optical depth.
For instance, the SED of stellar emission with $\eta =1000$ and
$\tau_V =1$ can be reproduced by the model with $\eta =100$ and $\tau_V$
slightly larger than 1.

\subsection {The Effect of System Size}

Figures 15a and 15b show $T_{eq}(r)$ and the SEDs for various scaling
factors $R$, respectively.
Note that the dust mass varies with $R$ for constant $\tau_V$.
The SED of absorbed stellar emission is independent of
the physical size of a system, provided that $\tau_V$ is kept constant.
Since $M_*$ does not vary, $T_{eq}(r)$ rises as a whole with decreasing $R$
due to the increase of energy density which leads to the shift of peak
of dust emission toward shorter wavelength.
The distribution of the source function remains the same, since both
distributions of stars and dust depend on $R$ in the same way.
The scaling relation for absolute value of the source
function $S^{(1)}$ is given
by $S^{(1)} \propto \varepsilon^* / n \sigma^e$, where
$\varepsilon^* \propto R^{-3}$. 
Since $\tau \propto n \sigma^e R$, we get 
finally $S^{(1)} \propto R^{-2}/\tau$. 
Here $\lambda$ dependence is omitted for clarity.
Since the resulting energy density of radiation field and
$S^{(1)}$ have the same dependence on $R$,
the integral in the r.h.s.\ of the equation (5)
is scaled by $R^{-2}$ when the mutual heating of dust is negligible.
That integral is proportional to $T_{eq}^{4+\theta}$,
if $\sigma^a_\lambda \propto \lambda^{-\theta}$.
Thus, $T_{eq}(r)$ is roughly scaled by $R^{-2/(4+\theta)}$.

The effect of $R$ on the peak wavelength of dust emission is
comparable to that of the age.
Once the age and $\tau_V$ are fixed, $R$ could be determined by the peak
wavelength of dust emission.
This point is important since the results of SED fitting
can be restricted by the size of the system derived from the observation.
Thus, it can help to determine the dust geometry
in the objects with $\tau_V \ll 1$, like
elliptical galaxies (see \S 4.3).

\subsection {The Effect of Extinction Curve}

Figures 16a and 16b show $T_{eq}(r)$ and the SEDs for the MW, LMC and SMC
extinction curve, respectively.
In order to enlarge differences among the SEDs, $\tau_V =30$
is adopted, instead of the standard value.
The systematic difference in $T_{eq}(r)$ should be noted, due to the
different strength of self-absorption indicated by the depth of
the silicate absorption feature at $9.7~ \mu$m.
The resulting shift of the peak of dust emission, however, is negligible in
comparison with those due to age or $R$.
Although the same $\tau_V$ is adopted for the cases of all
extinction curves, the resulting luminosity at V band is different
because of the different amount of scattered light.

\subsection{The Fraction of Absorbed Energy}

We introduce the energy fraction absorbed by dust, $\Delta E/E$.
The resulting SEDs are integrated from $91.2~ nm$ to $3~ \mu$m
to calculate the amount of unabsorbed energy, $E_*$, and
then $\Delta E/E$ is calculated by using the total energy, $E$, and
$\Delta E = E-E_*$.
In Figure 17 and Table 4, $\Delta E/E$ is
presented for various ages, $\tau_V$, $\eta$, and extinction curves.
Note that $\Delta E/E$ is independent of $R$ and is
equal to the energy fraction emitted by dust.
The dust distributed with larger $\eta$ absorbs the energy more
efficiently. This is due to an increase of the effective amount of dust
surrounding the concentrating stars.
For the constant value of $\tau_V$ and $\eta$, $\Delta E/E$
is always larger in younger system, since the absorbed UV photons
are more abundant.
The dependence on the extinction curve is negligible except for the
cases of youngest age 10 Myr and $\tau_V \le 1$.
$\Delta E/E$ increases rapidly as a function of $\tau_V$ when $\tau_V
\le 1$.
This strong dependence becomes less pronounced for an old system.
When $\tau_V \ge 1$, $\Delta E/E$ becomes more sensitive to the age
rather than $\tau_V$, especially for young system.

\subsection {Summary of Sensitivity Analysis}

We summarize the main results of our analysis:
1) the geometry of starburst galaxies should be described by 
$\eta \gtrsim 100$,
since the SED of absorbed stellar emission 
cannot be red enough to reproduce observed SEDs for $\eta \ll 100$. 
2) the UV-NIR SED is most sensitive to the stellar age and
optical depth $\tau_V$ when $\eta \gtsim 100$,
3) the FIR excess defined as $L_{FIR}/L_H$ is
sensitive to the age but not to $\tau_V$
when $\tau_V \ge 1$, which makes it useful for estimating the age of
system independently of $\tau_V$,
4) the dust emission reaches maximum at the FIR-submm region,
and the peak wavelength of dust emission depends mainly on the age and $R$.
From this study, we conclude that one can roughly estimate
the stellar age from $L_{FIR}/L_H$, $\tau_V$ from the
shape of the UV-NIR SED, and $R$ from the peak wavelength of dust emission
The depth of silicate absorption feature at 9.7 $\mu$m is useful to
constrain the type of extinction curve.
By using the resulting $R$ and the total stellar mass $M_*$ derived from the
normalisation of SED, the size of stellar system is calculated, 
which can be compared
with the observed size of the system to disentangle the degeneracy among $R$,
$\tau_V$ and $\eta$ for the optically thin case (see \S 4.3).

\section {Comparison with Observations}

The best-fitting SED model to observed data has been sought by trial 
and error, using the results of sensitivity analysis given in 
the previous section. The values of $\chi^2$ are calculated for each fit. 
Although, in the SED fitting procedure, data at the MIR wavelength region are
used to constrain $\tau_V$ and to choose the plausible extinction curve,
we exclude these data from the calculation of $\chi^2$. 
This is because the SED at
this wavelength region is very sensitive to the adopted optical properties
of silicate grains and PAHs, which are still a matter of debate, and
therefore cannot be used to judge the goodness of the fit.
For each galaxy, we show not only the best-fitting model, but also the
near-fitting models, which are helpful to infer the accuracy of
derived model parameters.

As it is shown in \S 3.3, $\eta$
should be larger than 100 to reproduce the strongly attenuated
SEDs of starburst galaxies,
although it is difficult to constrain the precise value of $\eta$
from the SED alone.
Here, we adopt $\eta=1000$ for starburst galaxies, as a result of which
the dust geometry is very close to the simple homogeneous distribution.
The model parameters to be determined
by observed SEDs are the age of stellar populations $t$,
optical depth in $V$-band, $\tau_V$, and the size scaling factor $R$.
For an elliptical galaxy NGC2768, the SED from UV to NIR is reproduced by
a galactic wind model, following Kodama \& Arimoto (1997), and then the
observed SED of dust emission is used to constrain $\tau_V$, $\eta$
and $R$. Thus, actually we fit SEDs of both starburst and elliptical
galaxies by employing three free parameters.
The resulting parameters for the best-fitting models and the derived
quantities are tabulated in Tables 5 and 6, respectively.

\subsection {Arp220}

Arp220 is the nearest ultraluminous infrared galaxy
at a distance 77 Mpc away from the Milky Way (Soifer et al. 1987).
Recent observations by ISO suggest that the source of dust heating
is nuclear starburst rather than the AGN (Lutz et al. 1996;
Lutz, Veilleux \& Genzel 1999).
Wynn-Williams \& Becklin (1993) derived the "compactness" of MIR
emission of Arp220. The compactness is defined
by the ratio of fluxes observed in two different apertures,
$5^{\prime \prime}.7$ and $100^{\prime \prime}$.
They show that about 99\% of MIR emission comes from the region within
$5^{\prime \prime}.7$ aperture.
Furthermore, Soifer et al. (1999) observed Arp220 with the Keck II
telescope to find that the 24.5 $\mu$m flux in the $4''$ diameter is
consistent with the flux density at 25 $\mu$m obtained by IRAS in
a rectangular beam, $\sim 1' \times 5'$.
Thus, MIR emission of Arp220 is concentrated in the central region.
Here, we confront our models to the central area  within
$5^{\prime \prime}$ aperture, which corresponds to $\sim 2.0$ kpc.

The photometric data are taken from Sanders et al. (1988; B, g, r, i),
Carico et al. (1990; J, H, K, L),
Smith, Aitken \& Roche (1989; MIR), Klaas et al. (1997; MIR to FIR),
NED (NASA/IPAC Extragalactic Database; IRAS bands) and
Rigopoulou, Lowrence \& Rowan-Robinson (1996; submm).
Recently, Goldader et al. (2002) presented the far-UV (1457 \AA)
and near-UV (2364 \AA) observations with HST/STIS.
As noted by the authors, these photometric data are problematic
because the large angular size of Arp220 is enough to fill
the entire field of view of STIS, which results in
the highly uncertain flux level of sky.
Therefore, these data are not used for the SED fitting,
although we show them in the related figures.

In Figures 18a--d, we show 9 SED models in total,
consisting of the best-fitting model
and 8 near-fitting models. The parameters for these models are tabulated
in Table 7 along with the value of reduced $\chi^2$. 
The best-fitting model is indicated by solid lines in all the four figures. 
It is found that the SMC extinction curve is required, in order to fit
simultaneously the strong silicate absorption feature at $9.7~ \mu$m and
the UV-NIR SED. Both the models with MW and LMC type extinction curve
overestimate the MIR fluxes, when the UV-NIR SED is well fitted. Therefore,
we use the SMC extinction curve for all the nine models.

The best-fitting model is giving by $t$=30 Myr, $\tau_V=40$ and 
$R=0.5$ with the SMC extinction curve, from which the bolometric luminosity
$L_{bol} = 1.4\ 10^{12} L_\odot$, the total stellar mass
$M_*=4.2\ 10^{10} M_\odot$ and 
the total dust mass $M_D=4.8\ 10^8 M_\odot$ are derived. 
The resulting $R=0.5$ corresponds to $r_t = 3.2$ kpc; thus,
roughly consistent with the adopted aperture size ($\sim$ 2.0 kpc). Note that,
in order to compare the size of starburst region, we should
be more precise for the definition of starburst region; for example,
the effective radius observed at some wavelength should be compared
with the effective radius of the model at the same wavelength.
However, such a detailed comparison of the geometry is not our main
purpose of this paper.

The derived value of reduced $\chi^2$ is relatively large even for the
best-fitting model. In the case of Arp220 and also M82, a significant
contribution to $\chi^2$ comes from only few most deviating photometric
data, e.g. at 3.7 $\mu$m for Arp220 and at 400 $\mu$m for M82. 
We would like to stress that the results of further fine-tuning 
of the SEDs parameters strongly depend on unknown systematic errors of 
observations performed with various telescopes, detectors and slightly 
different aperture sizes over a wide wavelength 
range. Indeed, genuine systematic uncertainties on data are expected. 
It is virtually impossible to obtain the exactly aperture-matched 
photometric data from the measurements performed with different aperture 
sizes and beam profiles for starburst regions whose surface brightness 
profile depends on the observed wavelengths (e.g. Scoville et al. 2000). 
Also, note that starburst regions are surrounded by underlying stellar 
populations with unclear boundaries. 
Furthermore, the absolute values of $\chi^2$ depend on the accuracy of 
statistical errors 
which were estimated quite inhomogeneously. Therefore, 
we refer the value of $\chi^2$ only to compare the goodness of 
fit among the reference models. 

The near-fitting models in Figure 18a have the same model parameters
as the best-fitting model, but age. Similarly, the different $\tau_V$
and $R$ are adopted for the near-fitting models in Figures 18b and 18c,
respectively.
The effects of age, $\tau_V$ and $R$ are
somewhat degenerate in this parameter range, especially due to
large $\tau_V$; i.e. 1) the variations of optical-NIR SED
due to the age effects are very similar to those due to $\tau_V$,
2) the average temperature of dust is changed by both the age and
$R$ effects.
On the other hand,
the UV luminosity is less sensitive in the range of age, $t$=15 -- 60 Myr.
This is because the bolometric luminosity is dominated by the UV
luminosity, intrinsically. Note that, in this case, the bolometric
luminosity is outlined by the FIR observations.

In Figure 18d, we show the best fitting model together with the models of
10 Myr and 300 Myr. Despite the large difference in age, the
optical-NIR SEDs become similar by adjusting both $\tau_V$ and $R$.
Therefore, the age of stellar populations is to be conservatively
determined in the range of 10 -- 300
Myr, when there are no constraints on $\tau_V$ and $R$
from the other observations.

Once the optical depth is determined independently of the SED,
the age can be constrained in the range of 15 -- 60 Myr as shown in Figure 18a.
In our model, $\tau_V$ is well restricted by the silicate absorption
feature when $\tau_V \gtrsim 10$. For Arp220, the best-fitting is achieved with
$\tau_V =40$ for $t=30$ Myr.
From the spectroscopic study with ISO, the extinction of
$\tau_V \approx 45$ was derived for Arp220 (Sturm et al. 1996; Genzel
et al. 1998).
The observations of radio recombination lines result in the similar
optical depth (Anantharamaiah et al. 2000).
These observational results seem to be consistent with our result.
However, it is misleading to compare these values directly, since
they are derived by  assuming the MW extinction curve.
Therefore, we compare optical depth at the NIR region, where hydrogen
recombination lines were observed by ISO to derive the extinction.
In our model for the SMC type extinction curve, $\tau_V =40$ corresponds to the
optical depth at K band, $\tau_K=2$, while the observed value is $\tau_K=4.5$.
This difference by a factor of 2 is not serious, since the observed optical
depths are inferred from the emission lines of ionized gas, whose distribution
could be different from that of dust. Calzetti, Kinney \&
Storchi-Bergmann (1994) suggest that the discrepancy by a factor of 2 is
rather typical for starburst galaxies. Considering these results,
we suggest that the constraint on age is rather tighter than the
conservative case noted above.

Among 9 models, only the oldest model (300 Myr) shown in Figure 18d is
marginally consistent with far-UV data, but overestimates near-UV data.
Although these data
are less reliable as noted above, the observation by HST/STIS actually
suggests very low fluxes at these wavelength,
even possible uncertainty of the sky level is taken into account. 
Such a low flux at the far-UV may be due to either
the selective extinction around massive stars or to unexpectedly large
anomaly of the extinction curve. When massive and hot stars tend to be
embedded in the clouds whose optical depth is considerably
larger than the average
which we derive, the far-UV light dominated by these stars can easily
be overestimated in our model. As for the extinction curve, it seems neither
MW, LMC nor SMC extinction curves could be applicable to starburst
regions, especially in the far-UV region in which very small
grains play dominant role, since
the highly energetic environments probably affect the size distribution
of dust grains. In order to reconcile the best-fitting model with
the observed far- and near-UV fluxes, the extinction should be extremely
effective at these wavelengths by an order of magnitude comparing with
the SMC extinction curve. These possibilities should be
investigated carefully in the future with reliable observations. 

\subsection{M82}

M82 is a nearby starburst galaxy at a distance of 3.3 Mpc.
This galaxy shows remarkable activities including the disturbed dusty
appearance and evidence for explosive star formation.
This may be caused by an interaction with
a neighbouring spiral galaxy M81 located at $\sim 36$ kpc away
in projection (Yun, Ho \& Lo 1993; Ichikawa et al. 1995).
The centre of this galaxy exhibits strong IR emission extending
$\sim 30^{\prime \prime}$ (Telesco \& Harper 1980) which coincides with
the effective radius of the bulge component (Ichikawa et al. 1995).
Thus, we confront our models to the SED of central starburst region within
$\sim 30^{\prime \prime}$ ($\sim$ 0.5 kpc).

The photometric data are taken from Johnson
(1966; U, B, V, R, I), Ichikawa et al. (1995, J, H, K),
Rice et al. (1988; IRAS bands), Telesco \& Harper (1980; FIR),
Jaffe, Becklin \& Hildebrand (1984; submm), and Elias et al. (1978; submm).
The spectrum from 2.4 $\mu$m to 45 $\mu$m is obtained with
the similar aperture size by the Short Wavelength Spectrometer (SWS)
on board the ISO (Sturm et al. 2000). From the ISO-SWS data,
we adopt the fluxes at 5, 16, 20, 30, 40 $\mu$m
for the calculation of $\chi^2$, at which the continuum emission
dominates.

In a similar manner to the case of Arp220,
we show the best-fitting model and near-fitting models in Figure 19a-d.
The parameters for these models are tabulated in Table 8, along with
the resulting values of reduced $\chi^2$. 
The best fit is achieved with $t$=30 Myr, $\tau_V$=8.5 and $R$=1.0.
We choose the LMC extinction curve for this galaxy,
judging from the depth of silicate absorption feature and the goodness
of fit to the optical-NIR SED.
Using this best-fitting model, we obtain
$L_{bol}=3.7\ 10^{10} L_\odot$, $M_*=1.1\ 10^{9} M_\odot$ and
$M_D=9.0\ 10^{6} M_\odot$.
The resulting $R$ corresponds to $r_t$=1.0 kpc, which is larger than
the observed size of starburst region by a factor of $\sim$4.

The behaviours of SED when changing $t$, $\tau_V$ and $R$ are almost
the same as for the case of Arp220. The notable difference is found for
the change of $\tau_V$, due to the smaller value of $\tau_V$ for M82;
i.e. the difference in $\tau_V$ affects
only the wavelength region shorter than $\sim 2\mu$m.
For a given age, the uncertainty of $\tau_V$ is at most
$\Delta \tau_V \sim 2$ as shown in Figure 19b,
while $\Delta \tau_V \sim 10$ for Arp220.
In the case of no restriction for both age and $\tau_V$, the
conservative range of age is 10 -- 100 Myr as shown in Figure 19d,
which is tighter than in the
case of Arp220 due to the smaller value of $\tau_V$.

The observed flux ratio of hydrogen recombination lines (Pa$\beta$/Br$\gamma$)
indicates $\tau_K \simeq 1.0$ in the central region (Satyapal et al. 1995).
We derive $\tau_K=0.5$ applying the LMC type extinction curve,
and thus find again a factor 2 discrepancy between the optical depth
derived from the stellar continuum and the line ratios.

In our model, the optical depth at 9.7$\mu$m is 1.1 for $\tau_V=8.5$ with
the LMC extinction curve, and therefore the silicate absorption is
important.
On the contrary, Sturm et al. (2000) concluded that no evidence for the strong
$9.7~ \mu$m absorption feature is found.
They compared the spectrum of M82 with that
of galactic reflection nebula NGC7023 which is expected to be free from the
foreground extinction.
There is no significant difference
in the spectra, except for $\lambda > 12~ \mu$m.
They succeeded to reproduce the spectrum of M82 up to
$\lambda \sim 12~ \mu$m with the scaled spectrum of NGC7023 plus a
power-law continuum which starts at $8.5~ \mu$m.
However, the solution obtained by their model may not be unique, since
the effect of self-absorption could be compensated by the assumed
flux of power-law continuum, which is a free parameter in their case,
but constrained by the other regions of SED in our model.

Thus, it is found that both the SEDs of M82 and Arp220 can be explained by
the stellar populations of the same age.
We note that, although the derived optical
depth is different by a factor of $\sim$5, Arp220 and M82 have the similar
mass ratio of dust to stars, $\sim 0.01$.
This also suggests that both Arp220 and M82 are at the similar
evolutionary phase as a starburst galaxy.
Therefore, the large difference between their SEDs
is mainly caused by the difference in the geometry; i.e. the starburst
region of Arp220 is much more compact than that of M82. This argument
is consistent with the obtained values of $R$, which are restricted
mainly by the peak wavelength of dust emission; i.e. the average
temperature of dust, in other words, the energy density of radiation
field and therefore the stellar density.

\subsection {HR10}

HR10 at $z=1.44$ is one of famous EROs having
$R-K > 6$ in the observed frame (Graham \& Dey 1996; Cimatti et al. 1998).
In general, extremely red colours are attributed either to very old stellar
populations or to young ones heavily obscured by dust
(Graham \& Dey 1996).
The latter case applies for HR10,
because of conspicuous dust emission
(Dey et al. 1999). The photometric data are all taken from Dey et al. (1999),
except for ISO data (Elbaz et al. 2002).
For this distant galaxy,
a precise choice of aperture for starburst region is difficult, and
therefore underlying stellar populations could contribute to the
observed fluxes. Dey et al. (1999) note that optical SEDs (in rest frame)
is similar to that of nearby ULIRGs rather than that of normal spiral
galaxies which is expected for underlying stellar populations in
gas-rich galaxies, like HR10. Therefore, we assume that observed fluxes
are dominated by starburst stellar populations.

In Figure 20, we show the best-fitting model with the age of 300 Myr,
and the near-fitting models with the age of 100 Myr and 1 Gyr.
We choose the MW extinction curve which gives better fit to rest
frame UV data.
The nearest fits are sought by fitting data excluding the MIR data,
which are also excluded in the calculation of reduced $\chi^2$
(see Table 9), like in the cases of Arp220 and M82. 

For the best-fitting model with $t=300$ Myr, $\tau_V=7$ and $R$=1.0,
we derive $M_*= 3.3\ 10^{11} M_\odot$ and $L_{bol}= 1.4\ 10^{12} L_\odot$.
The absorbed energy corresponds to $\sim 90$\% of the total one.
The dust mass $M_D = 1.2\ 10^9 M_\odot$ is larger than that of
the Arp220, despite a smaller $\tau_V$.
The resulting $R$ corresponds to $r_t=$18.2 kpc, while
the resolved size of this galaxy is $\sim 0^{\prime \prime}.9$
(Dey et al. 1999) corresponding to $\sim 5 h^{-1}_{75}$ kpc.

For HR10, the acceptable range of age is systematically older than
Arp220 and M82. The very red SED of HR10 at UV-optical cannot be
reproduced by the model with $t<100$ Myr.
The resulting mass ratio of dust to stars
is smaller than that of Arp220 and M82 by a factor of 2.
This is consistent with the old age of HR10 as a starburst; i.e.
gas, the fuel of starburst, decreases with aging, while
stellar mass increases as a result of star formation.

The SED without MIR data can be reproduced with the wide
range of starburst age (100 Myr -- 1 Gyr) as shown in the figure.
However, the oldest model is strongly rejected by MIR data.
Note that,
for most of starburst galaxies at high redshifts like those found by
the submillimetre common-user bolometer array (SCUBA) survey, MIR data are
not available, yet.
Also, observations at rest frame NIR and/or
at the emission peak of dust are extremely important for the tight constraint
on the age of high-$z$ starburst galaxies.

\subsection {NGC2768}

NGC2768, an elliptical galaxy (E6) at a distance of 21.5 Mpc,
is one of a few ellipticals that were detected in submm wavelength
(Wiklind \& Henkel 1995).
Photometric data are taken from Longo, Capaccioli \& Ceriello (1991; far-UV),
RC3 catalogue (de Vaucouleurs et al. 1991; UBV), Frogel et al. (1978; JHK),
and Wiklind \& Henkel (1995; submm). All data are reduced with 
the aperture (A) of total diameter (D) of the galaxy, i.e. $A/D$=1. 
Although Frogel et al. (1978) gave JHK data for $A/D=1$,
the aperture correction they adopted is not consistent with the
optical photometry. Therefore, we have adjusted JHK data to the system of
the optical data. In the far-UV bands, only the central part of galaxy
was observed by the IUE satellite (Longo et al.\ 1991).
Since the far-UV observations are performed in rectangular apertures
$10^{\prime \prime} \times 20^{\prime \prime}$,
we convert them to the effective circular apertures.
We have to note that such procedure is rather uncertain unless a true
far-UV photometric profile of the galaxy is known.

Since it is obvious that the stellar population in an elliptical galaxy
is by no means the SSP, we calculate the unabsorbed SED of NGC2768 by using
a galactic wind model that takes into account a realistic star formation
history in a galaxy (Arimoto \& Yoshii 1987; Kodama \& Arimoto 1997).
Model parameters are the same as Kodama \& Arimoto (1997)
who explained the empirical colour-magnitude relation of ellipticals in
Coma cluster as a sequence of mean stellar metallicity; i.e.
$t=12$ Gyr, $t_{SF}=0.1$ Gyr, $t_{in}=0.1$ Gyr and $x=1.1$, where $t$ is
a galactic age, $t_{SF}$ is a time scale of star formation,
$t_{in}$ is a time scale of gas infall, and $x$ is a
slope of the Salpeter-like IMF for the stellar mass interval
$0.10 \le m/M_{\odot} \le 60$. NGC2768 has $M_V=-21.8$ mag, from which
we derive $t_{gw} = 0.24$ Gyr for an epoch of galactic 
wind\footnote{Kodama \& Arimoto (1997) give different wind epoch,
since they adopted slightly steeper IMF with $x=1.2$}.

Figure 21 shows the best-fitting models for NGC2768.
Best fits are achieved for
$(\tau_V,\eta,R)=(0.3,1,12)$ and $(0.035,1000,0.7)$ as tabulated
in Table 10 with the value of reduced $\chi^2$. 
Although we choose the MW extinction curve, the resulting SEDs are not
sensitive to the extinction curve, since the amount of reddening
required is very small.
Obviously, the fit is not unique as the value of $\chi^2$ suggests,
because the SEDs from optical to NIR 
are insensitive to dust in the case of $\tau_V < 1$.
The degeneracy can be resolved if one takes into account a observed size of
the galaxy. The effective radius of NGC2768 is
$r_e=76^{\prime \prime}.5$ which corresponds
to $\sim 8$ kpc (Peletier et al. 1990).
Using the ratio $r_e/r_c \sim 10$ given by the adopted density
distribution of stars, we obtain $r_e=0.88$ and 11.0 kpc for $R=1.0$
and 12, respectively; each value of $R$ is converted to the real
size of galaxy by using $M_*= 2.2\ 10^{11} M_\odot$ for
$L_{bol}= 9.0\ 10^{10} L_\odot$.
Thus, the model with $R=12$ and $\eta=1$ fits better to the observation, giving
$M_D=2.9\ 10^6 M_\odot$.

If all the dust in NGC2768 comes from stars via stellar wind, the
amount of dust may be given as $M_D=\delta Z M_g$, where $\delta$ is
dust-to-heavy element ratio, and $M_g$ is the mass of gas.
The abundance of gas should be similar to the mean stellar metallicity of a galaxy.
Our galactic wind model fitted for NGC2768 gives [Fe/H]=$-0.09$, or
$\sim 0.80 Z_{\odot}$; thus $Z=0.016$.
After an epoch of galactic wind, the gas has been continuously
supplied to the interstellar space from evolving stars, and the amount
of gas should become as large as $\sim 20$\% of total stellar mass
(Arimoto 1989). This gives roughly $M_g \simeq 0.2 \times M_* =
4.3\ 10^{10} M_{\odot}$. Therefore, if the dust has always been kept in
a galaxy, its amount should be as large as $M_D \sim 2.7\ 10^8 M_{\odot}$
for $\delta = 0.4$. Then the average mass lose rate of dust during 12 Gyr is $\sim 0.02$
$M_{\odot}$ yr$^{-1}$. The amount of dust we derive from the SED fitting
is $2.9\ 10^6 M_{\odot}$, which is much smaller than the value expected and
corresponds to the amount of dust accumulated during the last
$\sim 0.1$ Gyr. Obviously, most of dust ever ejected from stars was lost
to the extragalactic space and/or evaporated.

\section{Discussion}
\subsection{Star formation history of starbursts}
In our model, the intrinsic stellar SEDs of the SSP are used.
Since galaxies are composed of composite stellar populations, the
elapsed time from the onset of starburst, or simply absolute age of
starburst can be derived only if the star formation history is known.
In the model of Efstathiou et al (2000), the star formation history
was parameterized with a time scale for an exponential decay of
star formation rate, which is estimated from the SED.
For M82, they required two bursts of star formation with different
time scales (2 Myr and 6 Myr).
They derived the starburst ages of 26.5 and 16.5 Myr for each starburst
event, respectively.
On the other hand, in our model, the SED of M82 can be reproduced
with a single starburst population; i.e. no need to assume multiple
starburst events, although we derive the age very similar 
to that derived by Efstathiou et al. (2000). 
Note that Silva et al. (1998) assumed different star formation
history of starbursts as well as the starburst age and applied
to M82.
This means that the derived star formation history depends on
the model assumptions, strongly.
Since, at least, these models can reproduce the
observed SEDs of M82, we clearly need the other observational constraints,
in order to estimate the star formation history of starbursts.
The efficiency of star formation which is important clue to 
the star formation history can be inferred from 
the ratio of the indicator of
star formation rate, such as the bolometric luminosity, 
to the total mass of gas in the starburst region (Sanders et al. 1988).
We suggest that an evolutionary SED model in which the chemical
evolution is taken into account, could be helpful to derive the absolute
age of starburst, when the observed SED and gas mass
are reproduced with the evolutionary SED model, simultaneously.

\subsection{Implications for high redshift starbursts}
Recently, various galaxy populations have been found as candidates of
starburst galaxies at high redshifts. These populations can be
classified by the wavelength ranges in which galaxy populations are defined;
one class consists of galaxies whose characteristics of
stellar emission are used as a criterion, such as Lyman-break galaxies
and EROs, and for another class dust emission is used, like
galaxies detected by ISO and/or SCUBA.
It is important to investigate
the relationship between these two classes of galaxy population,
in order to reveal the nature of high-$z$ starburst galaxies.
Lyman-break galaxies are too faint at submm wavelengths to be
detected with the currently available telescopes (e.g. Ouchi et al. 1999).
EROs have been expected as optical counterparts for SCUBA galaxies,
since their extreme red colours can be explained by the significant
dust extinction.
Although this prospect seems to be confirmed with the submm detection of
a prototypical ERO HR10,
Cimatti et al. (2002) however found that the star formation rates
of star-forming EROs
with $R-K>5$ are an order of magnitude less than those of SCUBA galaxies.
This result is consistent with
the negative SCUBA detection for $R-K$ EROs.
Mohan et al. (2002) detected none of 27 EROs
defined with $R-K>5$ or $I-K>5$, but one, in
the submm wavelengths.
Thus, it is not yet clear even whether there is
a representative optical/NIR galaxy population for SCUBA
galaxies.

Great opportunities to study statistical properties of
high-$z$ starburst galaxies will be
provided by the future infrared space missions, such as
SIRTF-SWIRE (Franceschini et al. 2002)
and ASTRO-F (Nakagawa et al. 2001).
In both missions, the wide range of wavelength
from NIR to FIR will be covered, as a result of which both stellar
and dust emission of high-$z$ starbursts
can be observed. Therefore, these missions
will be very important to investigate the relationship
between the galaxy classes characterized by stellar and dust emissions.
Our model is optimal for this analysis, since
all regions of SED are physically interrelated, as it was demonstrated for the
nearby starburst galaxies. Furthermore, starburst ages, stellar masses
and optical depths could be derived for statistical sample of
high-$z$ starburst galaxies, as a result of model application
to the data obtained with both missions.
These results should provide important clues on the process of
galaxy formation.

\section{Conclusions}
We have developed a code which solves the equation of radiative
transfer by assuming spherical symmetry for arbitrary radial distributions
of stars and dust.
Isotropic multiple scattering is assumed and the self-absorption of
re-emitted energy from dust is fully taken into account.
The adopted dust model takes into account graphite and
silicate grains, and PAHs.
We successfully reproduce the MW, LMC and SMC type extinction curves as well
as the spectra of cirrus emission in the MW.

We adopt the geometry where stars and dust are distributed by the 
King model.
The SEDs of starburst galaxies are reproduced well
with the core radius ratio of dust to stars $\eta \ge 100$;
thus, dust distributes nearly uniformly and surrounds
centrally concentrating stars.
The UV-NIR SED is most sensitive to the age and optical depth $\tau_V$.
The FIR excess defined by $L_{FIR}/L_H$ depends mainly on
age, and is rather insensitive to $\tau_V$ when $\tau_V \geq 1$, which
allows rough estimate of starburst age by using $L_{FIR}/L_H$.
The peak wavelength of dust emission is in the FIR-submm region,
and depends mainly on the age and size scaling factor $R$.
We show that the extinction curve in starburst galaxies is close to that
in the SMC (Arp220) or in the LMC (M82), judging from the MIR features
when the other parts of SED are reproduced.
Therefore, the non-MW type extinction curve should be considered,
in order to model the SEDs of starburst galaxies.

UV-submm SEDs of starburst galaxies, Arp220, M82 and HR10, can be
reproduced consistently with the reddened starburst stellar populations.
The derived age of stellar population in
the starburst region of Arp220 is 30 Myr for
an instantaneous starburst (SSP model).
Although the same age is derived for M82, the obtained optical depth is
significantly different ($\tau_V=8.5$ for M82, and 40 for Arp220).
The restriction to the starburst age of M82 is tighter
than that of Arp220, due to the smaller optical depth.
For prototypical star-forming ERO HR10, we derive systematically
older age (300 Myr) than Arp220 and M82, which is required to reproduce
the extremely red colour at optical-NIR region. This difference in
evolutionary phase as a starburst 

is also suggested by the smaller mass ratio
of dust to stars in HR10 than that of Arp220 and M82.

The optically thin limit is studied in case of the elliptical galaxy NGC2768,
where $\tau_V$, galaxy size, and $\eta$ are degenerate.
Determination of the galactic size from observation is required
to disentangle this degeneracy. From the observed size and SED of
this galaxy, the distribution of dust is found to be similar to that
of stars.

\vspace{1pc}\par
T.T. would like to thank T. Shibazaki and late R. Hoshi for their
continuous support and fruitful discussions.
V.V. are grateful to National Astronomical Observatory of Japan
for hospitality during his stay as the guest professor during
which a part of work was done.
This work was financially supported in part by a Grant-in-Aid for
Scientific Research (No. 1164032) by the Japanese
Ministry of Education, Culture, Sports and Science, and
by a Grant of the Lithuanian State Science and Studies Foundation.
This research has made use of the NASA/IPAC Extragalactic Database
(NED) which is operated by the Jet Propulsion Laboratory,
California Institute of Technology, under contract with the
National Aeronautics and Space Administration.

\newpage

\section*{References}
\re
Anantharamaiah K.R., Vaillefond F., Mohan N.R., Goss W.M., Zhao J.H.\ 2000,
ApJ, 537, 613
\re
Arimoto N.\ 1989, in Evolutionary Phenomena in Galaxies, eds. J.E. Beckman,
B.E.J. Pagel (Cambridge University Press) p. 323
\re
Arimoto N., Yoshii Y.\ 1986, A\&A, 164, 260
\re
Arimoto N., Yoshii Y.\ 1987, A\&A, 173, 23
\re
Band D.L., Grindlay J.E.\ 1985, ApJ, 298, 128
\re
Bithell M., Rees M.J.\ 1990, MNRAS, 242, 570
\re
Boulanger F., Boissel P., Cesarsky D., Ryter C.\ 1998, A\&A, 339, 194
\re
Bressan A., Chiosi C., Fagotto F.\ 1994, ApJS, 94, 63
\re
Calzetti D., Kinney A.L., Storchi-Bergmann T.\ 1994, ApJ, 429, 582
\re
Carico D.P., Sanders D.B., Soifer B.T., Matthews K., Neugebauer G.\
1988, AJ, 95, 356
\re
Carico D.P., Sanders D.B., Soifer B.T., Matthews K., Neugebauer G.\
1990, AJ, 100, 70
\re
Chambers K.C., Charlot S.\ 1990, ApJ, 348, L1
\re
Chambers K.C., Miley G.K., van Breugel W.J.M.\ 1990, ApJ, 363, 21
\re
Chan K.-W., Roellig T.L., Onaka T., Mizutani M., Okumura K., Yamamura I.,
Tanab\'e T., Shibai H. et al.\ 2001, ApJ, 546, 273
\re
Cimatti A., Andreani P., R\" ottgering H., Tilanus R.\ 1998, Nature, 392, 895
\re
Cimatti A., Daddi E., Mignoli M., Pozzetti L., Renzini A., Zamorani G., Broadhurst T.,
Fontana A. et al.\  2002, A\&A, 381, L68
\re
Combes F., Boiss\' e P., Mazure A., Blanchard A.\ 1995, in Galaxies
and Cosmology (Springer-Verlag, Berlin Heidelberg) p. 96
\re
D\'esert F.-X., Boulanger F., Puget J.L.\ 1990, A\&A, 237, 215
\re
D\'esert F.-X., Boulanger F., Shore S.N.\ 1986, A\&A, 160, 295
\re
Devriendt J.E.G., Guiderdoni B., Sadat R.\ 1999, A\&A, 350, 381
\re
de Vaucouleurs G., de Vaucouleurs A., Corwin H.G.Jr., Buta R., Paturel
G.,
Fouque P.\ 1991, in The Third Reference Catalogue of Bright Galaxies
(Springer Verlag, New York) [RC3]
\re
Dey A., van Breugel W., Vacca W.D., Antonucci R.\ 1997, ApJ, 490, 698
\re
Dey A., Graham J.R., Ivison R.J., Smail I., Wright G.S., Liu M.C.\ 1999,
ApJ, 519, 610
\re
Draine B.T., Anderson N.\ 1985, ApJ, 292, 494
\re
Draine B.T., Lee H.M.\ 1984, ApJ, 285, 89
\re
Dwek E.\ 1986, ApJ, 302, 363
\re
Dwek E., Arendt R. G., Fixsen D. J., Sodroski T. J., Odegard N., Weiland J. L.,
Reach W. T., Hauser, M. G.\ et al.\ 1997, ApJ, 475, 565
\re
Eales S.A., Rawlings S., Dickinson M., Spinrad H., Hill G.J., Lacy M.\
ApJ, 1993, 409, 578
\re
Elbaz D., Flores H., Chanial P., Mirabel I.F., Sanders D., P.-A. Duc, Cesarsky C.J.,
Aussel H.\ 2002, A\&A, 381, L1
\re
Efstathiou A., Rowan-Robinson M., Siebenmorgen R.\ 2000, MNRAS, 313, 734
\re
Eisenhardt P., Dickinson M.\ 1992, ApJ, 399, L47
\re
Elias J.H., Ennis D.J., Gezari D.Y., Hauser M.G., Houck J.R.,
Lo K.Y., Matthews K., Nadeau D. et al.\ 1978, ApJ, 220, 25
\re
Fagotto F., Bressan A., Bertelli G., Chiosi C.\ 1994a, A\&AS, 104, 365
\re
Fagotto F., Bressan A., Bertelli G., Chiosi C.\ 1994b, A\&AS, 105, 29
\re
Fioc M., Rocca-Volmerange B.\ 1997, A\&A, 326, 950
\re
Franceschini A., Lonsdale C., SWIRE Co-Investigator Team\ 2002, astro-ph/0202463
\re
Frogel J.A., Persson S.E., Aaronson M., Matthews K.\ 1978, ApJ, 220, 75
\re
Genzel R., Lutz D., Sturm E., Egami E., Kunze D., Moorwood A.F.M.,
Rigopoulou D., Spoon H.W.W.\ et al.\ 1998, A\&A, 498, 579
\re
Goldader J.D., Joseph R.D., Doyon R., Sanders D.B.\ 1995, ApJ, 444, 97
\re
Goldader J.D., Meurer G., Heckman T.M., Seibert M., Sanders D.B., Calzetti D., Steidel C.C.\ 2002,
ApJ, 568, 651
\re
Gordon K.D., Calzetti D., Witt A.N.\ 1997, ApJ, 487, 625
\re
Graham J.R., Dey A.\ 1996, ApJ, 471, 720
\re
Graham J.R., Matthews K., Soifer B.T., Nelson J.E., Harrison W.,
Jernigan J.G., Lin S. et al.\ 1994, ApJ, 420, L5
\re
Guhathakurta P., Draine B.T.\ 1989, ApJ, 345, 230
\re
Gunn J.E., Stryker L.L.\ 1983, ApJS, 52, 121
\re
Ichikawa T., Yanagisawa K., Itoh N., Tarusawa K., van Driel W.,
Ueno M.\ 1995, AJ, 109, 2038
\re
Ivezi\' c \v Z., Groenewegen M.A.T., Men'shchikov A., Szczerba R.\
1997, MNRAS, 291, 121
\re
Jaffe D.T., Becklin E.E., Hildebrand R.H.\ 1984, ApJ, 285, L31
\re
Johnson H.L.\ 1966, ApJ, 143, 187
\re
Klaas U., Haas M., Heinrichsen I., Schulz B.\ 1997, A\&A, 325, L21
\re
Kobayashi Y., Sato S., Yamashita T., Shiba H.,Takami H.\ 1993, ApJ, 404, 94
\re
Kodama T., Arimoto N.\ 1997, A\&A, 320, 41
\re
Kurucz R.L.\ 1992, The stellar Population of Galaxies,
eds. B. Barbuy, A. Renzini (Dordrecht: Kluwer), p. 225
\re
Lacy M., Rawlings S.\ 1996, MNRAS, 280, 888
\re
Laor A., Draine B.T.\ 1993, ApJ, 402, 441
\re
L\' eger A., d'Hendecourt L., D\' efourneau D.\ 1989, A\&A, 216, 148
\re
Leitherer C., Heckman T.M.\ 1995, ApJS, 96, 9
\re
Lemke D., Mattila K., Lehtinen K., Laureijs R. J., Liljestrom T.,
L\' eger, A., Herbstmeier U.\ 1998, A\&A, 331, 742
\re
Lilly S.J.\ 1988, ApJ, 333, 161
\re
Longo G., Capaccioli M., Ceriello A.\ 1991, A\&AS, 90, 375
\re
Low F.J., Young E., Beintema D.A., Gautier T.N., Beichman C.A.,
Aumann H.H., Gillett F.C., Neugebauer G.\ et al.\ 1984, ApJ, 278, L14
\re
Lutz D., Genzel R., Sternberg A., Netzer H., Kunze D., Rigopoulou D.,
Strum E., Egami E. et al.\ 1996, A\&A, 315, L137
\re
Lutz D., Veilleux S., Genzel R.\ 1999, ApJ, 517, L13
\re
Mathis J.S.\ 1990, ARA\&A, 28, 37
\re
Mathis J.S., Mezger P.G., Panagia N.\ 1983, A\&A, 128, 212
\re
Mathis J.S., Rumpl W., Nordsieck K.H.\ 1977, ApJ, 217, 425
\re
Mazzei P., De Zotti G., Xu C.\ 1994, ApJ, 422, 81
\re
Meurer G.R., Heckman T.M., Lehnert M.D., Leitherer C., Lowenthal J.,
1997, AJ, 114, 54
\re
Meurer G.R., Heckman T.M., Leitherer C., Kinney A., Robert C.,
Garnett D.R.\ 1995, AJ, 110, 2665
\re
Mihalas D.\ 1978, Stellar Atmospheres (W.H. Freeman and Company,
San Francisco), ch. 2
\re
Mohan N.R., Cimatti A., Rottgering H.J.A., Andreani P., Severgnini P., Tilanus R.P.J., Carilli C.L.,
Stanford S.A.\ 2002, A\&A, 383, 440
\re
Nakagawa T.\ 2001, in The promise of Herschel Space Observatory, 
ed. G.L. Pilbratt, J. Cernicharo, A.M. Heras, T. Prusti, 
R. Harris, (Noordwijk ESA-SP 460), p.67 
\re
O'Connell R.W., Gallagher J.S., Hunter D.A.\ 1994, ApJ, 433, 65
\re
Omont A.\ 1986, A\&A, 164, 159
\re
Ouchi M., Yamada T., Kawai H., Ohta K.\ 1999, ApJ, 517, L19
\re
Pickles A.J.\ 1985, ApJS, 59, 33
\re
Pei Y.C.\ 1992, ApJ, 395, 130
\re
Peletier R.F., Davies R.L., Illingworth G.D., Davis L.E., Cawson M.\
1990, AJ, 100, 1091
\re
Puget J.L., L\' eger A.\ 1989, ARA\&A, 27, 161
\re
Rakos K.D., Maindl T.I., Schombert J.M.\ 1996, ApJ, 466, 122
\re
Rice W., Lonsdale C.J., Soifer B.T., Neugebauer G., Koplan E.L.,
Lloyd L.A., de Jong T., Habing H., J.\ 1988, ApJS, 68, 91
\re
Rieke G.H., Lebofsky M.J.\ 1985, ApJ, 288, 618
\re
Rigopoulou D., Lawrence A., Rowan-Robinson M.\ 1996, MNRAS, 278, 1049
\re
Rowan-Robinson M.\ 1980, ApJS, 44, 403
\re
Rowan-Robinson M., Efstathiou A.\ 1993, MNRAS, 263, 675
\re
Sanders D.B., Soifer B.T., Elias J.H., Madore B.F., Matthews K.,
Neugebauer G., Scoville N.Z.\ 1988, ApJ, 325, 74
\re
Satyapal S., Watson D.M., Pipher J.L., Forrest W.J., Coppenbarger D.,
Raines S.N., Libonate S., Piche F.\ et al.\ 1995, ApJ, 448, 611
\re
Schmitt H.R., Kinney A.L., Calzetti D., Bergmann T.S.\ 1997, AJ, 114, 592
\re
Scoville N.Z., Evans A.S., Thompson R., Reike M., Hines D.C.,
Low F.J., Dinshaw N., Surace J.A.\ et al.\ 2000, AJ, 119, 991
\re
Shioya Y., Taniguchi Y., Trentham N.\ 2001, MNRAS, 321, 11
\re
Siebenmorgen R., Kr\"ugel E.\ 1992, A\&A, 259, 614
\re
Silva L., Granato G.L., Bressan A., Danese L.\ 1998, ApJ, 509, 103
\re
Smith C.H., Aitken D.K., Roche P.F.\ 1989, MNRAS, 241, 425
\re
Soifer B.T., Sanders D.B., Madore B.F., Neugebauer G., Danielson D.E.,
Elias J.H., Lonsdale C.J., Rice W.L.\ 1987, ApJS, 320, 238
\re
Soifer B.T., Neugebauer G., Matthews K., Becklin E.E., Ressler M.,
Werner M.W., Weinberger A.J., Egami E.\ 1999, ApJ, 513, 207
\re
Sturm E., Lutz D., Genzel R., Sternberg A., Egami E., Kunze D.,
Rigopoulou D., Bauer O.H.\ et al.\ 1996, A\&A, 315, L133
\re
Sturm E., Lutz D., Tran D., Feuchtgruber H., Genzel R., Kunze D.,
 Moorwood A.F.M., Thornley M. D.\ 2000, A\&A, 358, 481
\re
Takagi T., Arimoto N., Vansevi\v cius V.\ 1999, ApJ, 523, 107
\re
Telesco C.M., Harper D.A.\ 1980, ApJ, 235, 392
\re
Tacconi-Garman L.E., Sternberg A., Eckart A.\ 1996, AJ, 112, 918
\re
Totani T., Takeuchi T.T.\ 2002, ApJ, 570, 470
\re
VandenBerg D.A., Hartwick F.D.A., Dawson P., Alexander D.R.\ 1983,
ApJ, 266, 747
\re
Verstraete L., L\'eger A.\ 1992, A\&A, 266, 513
\re
Vansevi\v cius V., Arimoto N., Kodaira K.\ 1997, ApJ, 474, 623
\re
Vassiliadis E., Wood P.R.\ 1993, ApJ, 413, 641
\re
Vassiliadis E., Wood P.R.\ 1994, ApJS, 92, 125
\re
Weingartner J.C., Draine B.T.\ 2001, ApJS, 134, 263
\re
Wiklind T., Henkel C.\ 1995, A\&A, 297, L71
\re
Windhorst R.A., Burstein D., Mathis D.F., Neuschaefer L.M., Bertola F.,
Buson L.M., Koo D.C., Matthews K. et al.\ 1991, ApJ, 380, 362
\re
Witt A.N., Thronson H.A., Capuano J.M.\ 1992, ApJ, 393, 611
\re
Wynn-Williams C.G., Becklin E.E.\ 1993, ApJ, 412, 535
\re
Yi S.\ 1996, Ph.D. Thesis, Yale University
\re
Yun M.S., Ho P.T.P., Lo K.Y.\ 1993, ApJ, 411, L17

\newpage

\begin{table*}[t]
\begin{center}
Table~1.\hspace{4pt} Parameters of PAH features\\
\vspace{6pt}
\begin{tabular*}{230pt}{@{\hspace{\tabcolsep}
\extracolsep{\fill}}p{3pc}cccc}
\hline\hline\\ [-20pt]
\multicolumn{1}{c}{$\lambda [\mu m]$} &
\multicolumn{1}{c}{$\nu_0 [cm^{-1}]$} &
\multicolumn{1}{c}{$\Gamma [cm^{-1}]$} &
\multicolumn{1}{c}{$A [10^{-25}cm^3$]} \\
\hline \\[-20pt]
3.3\dotfill & 3030 & 23.5 & 1.4 $N_H$  \\
6.2\dotfill  & 1610 & 36.5 & 0.7 $N_C$  \\
7.7\dotfill  & 1300 & 59.0 & 2.0 $N_C$  \\
8.6\dotfill  & 1160 & 34.5 & 1.2 $N_H$  \\
11.3\dotfill & 885  & 17.0 & 14.1 $N_H$  \\
12.7\dotfill & 787  & 32.5 & 11.1 $N_H$  \\
\hline \\
\end{tabular*}
\end{center}
\vspace{-15pt}
\par\noindent
Note. $N_C$ and $N_H$ are the number of carbon and hydrogen atoms in a
PAH molecule, respectively.\\
\end{table*}

\tabcolsep=1.0mm
\begin{table*}[t]
\begin{center}
Table~2.\hspace{4pt} The relative fractions of dust particles \\
\vspace{6pt}
\begin{tabular*}{300pt}{@{\hspace{\tabcolsep}
\extracolsep{\fill}}p{5pc}cccccc}
\hline\hline\\ [-10pt]
& $M_D/M_H$ & \multicolumn{2}{c}{Graphite} &
 \multicolumn{2}{c}{Silicate}    &
 PAH \\
\hline\\ [-20pt]
 & & BG & VSG &BG &VSG & \\
\hline\\ [-20pt]
MW \dotfill & 8.0 $10^{-3}$ & 0.37 & 0.08 & 0.43 & 0.07 & 0.05\\
LMC\dotfill & 3.6 $10^{-3}$ & 0.10 & 0.02 & 0.74 & 0.13 & 0.01\\
SMC\dotfill & 1.9 $10^{-3}$ & 0.04 & 0.009& 0.808 & 0.138 & 0.005\\
\hline \\
\end{tabular*}
\end{center}
\vspace{-15pt}
\par\noindent
Note. $M_D/M_H$ is the mass ratio of dust to hydrogen.
\end{table*}

\tabcolsep=1.0mm
\begin{table*}[t]
\begin{center}
Table~3.\hspace{4pt} Parameters of the standard model \\
\vspace{6pt}
\begin{tabular*}{300pt}{@{\hspace{\tabcolsep}
\extracolsep{\fill}}p{3pc}ccccc}
\hline\hline\\ [-10pt]
 Age & $\tau_V$ &  $R$ & $\eta$ & Extinction Curve \\
\hline\\ [-20pt]
10 Myr & 1.0  & 1.0 &  1000 & MW \\
\hline \\
\end{tabular*}
\end{center}
\end{table*}

\tabcolsep=1.0mm
\begin{table*}[t]
\begin{center}
Table~4.\hspace{4pt}Fraction (\%) of the absorbed energy \\
\end{center}
\vspace{6pt}
\begin{tabular*}{\textwidth}{@{\hspace{\tabcolsep}
\extracolsep{\fill}}p{3pc}cccccccccccc}
\hline\hline\\ [-10pt]
\multicolumn{13}{c}{MW Extinction Curve} \\
\hline \\ [-20pt]
 & \multicolumn{4}{c}{10 Myr}
 & \multicolumn{4}{c}{1 Gyr}
 & \multicolumn{4}{c}{10 Gyr}\\
\hline \\ [-20pt]
\multicolumn{1}{c}{$\tau_V$}    &
\multicolumn{1}{c}{1$^\dagger$} &
\multicolumn{1}{c}{10$^\dagger$}&
\multicolumn{1}{c}{100$^\dagger$} &
\multicolumn{1}{c}{1000$^\dagger$}&
\multicolumn{1}{c}{1} &
\multicolumn{1}{c}{10} &
\multicolumn{1}{c}{100} &
\multicolumn{1}{c}{1000} &
\multicolumn{1}{c}{1} &
\multicolumn{1}{c}{10} &
\multicolumn{1}{c}{100} &
\multicolumn{1}{c}{1000}  \\
\hline \\[-20pt]
0.5 \dotfill &6&26&45&49&4&11&18&20&4&9&14&15  \\
1.0 \dotfill &10&39&66&71&5&18&31&34&5&14&24&26 \\
2.0 \dotfill &15&52&83&87&8&29&49&53&7&24&40&44 \\
5.0 \dotfill &23&66&93&96&14&46&74&78&12&40&66&71 \\
10.0\dotfill &30&74&97&98&19&58&87&90&17&53&82&86 \\
\hline\\
\multicolumn{13}{c}{SMC Extinction Curve} \\
\hline \\ [-20pt]
 & \multicolumn{4}{c}{10 Myr}
 & \multicolumn{4}{c}{1 Gyr}
 & \multicolumn{4}{c}{10 Gyr}\\
\hline \\ [-20pt]
\multicolumn{1}{c}{$\tau_V$}    &
\multicolumn{1}{c}{1} &
\multicolumn{1}{c}{10} &
\multicolumn{1}{c}{100} &
\multicolumn{1}{c}{1000}&
\multicolumn{1}{c}{1} &
\multicolumn{1}{c}{10} &
\multicolumn{1}{c}{100} &
\multicolumn{1}{c}{1000} &
\multicolumn{1}{c}{1} &
\multicolumn{1}{c}{10} &
\multicolumn{1}{c}{100} &
\multicolumn{1}{c}{1000}  \\
\hline \\[-20pt]
0.5 \dotfill &8&32&55&59&4&11&19&20&4&9&14&15 \\
1.0 \dotfill &13&45&72&77&6&19&32&34&5&14&24&26 \\
2.0 \dotfill &18&56&84&88&8&29&49&53&7&23&40&43 \\
5.0 \dotfill &26&69&94&96&14&46&73&77&12&40&65&70 \\
10.0\dotfill &33&76&97&98&19&57&86&89&17&52&81&85 \\
\hline
\end{tabular*}
\vspace{6pt}
\par\noindent
$\dagger$ Core radius ratio, $\eta$
\end{table*}

\begin{table*}[t]
\begin{center}
Table~5.\hspace{4pt} Parameters of the best-fitting models\\
\vspace{6pt}
\begin{tabular*}{300pt}{@{\hspace{\tabcolsep}
\extracolsep{\fill}}p{3pc}ccccccc}
\hline\hline\\ [-10pt]
\multicolumn{1}{c}{Name} &
\multicolumn{1}{c}{$t$ [Gyr]} &
\multicolumn{1}{c}{$\tau_V$} &
\multicolumn{1}{c}{$R$} &
\multicolumn{1}{c}{$\eta$} &
\multicolumn{1}{c}{Extinction Curve} &
\multicolumn{1}{c}{Reduced $\chi^2$} \\
\hline \\[-20pt]
Arp220& 0.03 & 40 & 0.5 &1000 & SMC & 1.69\\
M82& 0.03 & 8.5 & 1.0 &1000 & LMC & 1.85\\
HR10 & 0.3 & 7.0 & 1.0 &1000 &  MW & 1.64 \\
NGC2768& 12$^a$ & 0.3 & 12.& 1 & MW$^a$ & 1.67\\
\hline \\
\end{tabular*}
\end{center}
\vspace{-16pt}
\par\noindent
$^a$ assumed parameters\\
\end{table*}

\begin{table*}[t]
\begin{center}
Table~6.\hspace{4pt} Derived galaxy luminosities and masses\\
\vspace{6pt}
\begin{tabular*}{300pt}{@{\hspace{\tabcolsep}
\extracolsep{\fill}}p{3pc}cccccccc}
\hline\hline\\ [-10pt]
\multicolumn{1}{c}{Name} &
\multicolumn{1}{c}{$M_*$} &
\multicolumn{1}{c}{$L_{bol}$} &
\multicolumn{1}{c}{$M_D$} \\
\multicolumn{1}{c}{} &
\multicolumn{1}{c}{[$10^{11} M_\odot$]} &
\multicolumn{1}{c}{[$10^{11} L_\odot$]} &
\multicolumn{1}{c}{[$10^{7} M_\odot$]}  \\
\hline \\[-20pt]
Arp220& 0.42 & 14. & 48.  \\
M82&  0.011 & 0.37 & 0.90  \\
HR10 &  3.3 & 14. & 120. \\
NGC2768&  2.2 & 0.9 & 0.29  \\
\hline \\
\end{tabular*}
\vspace{6pt}
\end{center}
\end{table*}

\begin{table*}[t]
\begin{center}
Table~7.\hspace{4pt} Model parameters for Arp220\\
\vspace{6pt}
\begin{tabular*}{300pt}{@{\hspace{\tabcolsep}
\extracolsep{\fill}}p{3pc}ccccccc}
\hline\hline\\ [-10pt]
\multicolumn{1}{c}{Name} &
\multicolumn{1}{c}{$t$ [Myr]} &
\multicolumn{1}{c}{$\tau_V$} &
\multicolumn{1}{c}{$R$} &
\multicolumn{1}{c}{$\eta$} &
\multicolumn{1}{c}{Extinction curve} &
\multicolumn{1}{c}{Reduced $\chi^2$} \\
\hline \\[-20pt]
model 1  & 30  & 40 & 0.5 & 1000 & SMC & 1.69\\
model 2  & 15  & 40 & 0.5 &1000  & SMC & 2.94\\
model 3  & 60  & 40 & 0.5 &1000  & SMC & 5.37\\
model 4  & 30  & 60 & 0.5 &1000  & SMC & 2.34\\
model 5  & 30  & 30 & 0.5 &1000  & SMC & 4.63\\
model 6  & 30  & 40 & 0.3 &1000  & SMC & 3.62\\
model 7  & 30  & 40 & 1.0 &1000  & SMC & 2.74\\
model 8  & 300  & 70 & 0.1 & 1000 & SMC & 3.06\\
model 9  & 10  & 30 & 1.0 &1000  & SMC & 3.98\\
\hline \\
\end{tabular*}
\end{center}
\end{table*}

\begin{table*}[t]
\begin{center}
Table~8.\hspace{4pt} Model parameters for M82\\
\vspace{6pt}
\begin{tabular*}{300pt}{@{\hspace{\tabcolsep}
\extracolsep{\fill}}p{3pc}ccccccc}
\hline\hline\\ [-10pt]
\multicolumn{1}{c}{Name} &
\multicolumn{1}{c}{$t$ [Myr]} &
\multicolumn{1}{c}{$\tau_V$} &
\multicolumn{1}{c}{$R$} &
\multicolumn{1}{c}{$\eta$} &
\multicolumn{1}{c}{Extinction curve} &
\multicolumn{1}{c}{Reduced $\chi^2$} \\
\hline \\[-20pt]
model 1  & 30  & 8.5 & 1.0 &1000 & LMC & 1.85\\
model 2  & 15  & 8.5 & 1.0 &1000 & LMC & 3.35\\
model 3  & 60  & 8.5 & 1.0 &1000 & LMC & 8.83\\
model 4  & 30  & 10. & 1.0 &1000 & LMC & 2.51\\
model 5  & 30  & 7.0 & 1.0 &1000 & LMC & 2.64\\
model 6  & 30  & 8.5 & 1.5 &1000 & LMC & 6.95\\
model 7  & 30  & 8.5 & 0.5 &1000 & LMC & 3.75\\
model 8  & 10  & 8.0 & 0.1 &1000 & LMC & 2.72\\
model 9  & 100  & 10. & 1.0 &1000 & LMC & 3.72\\
\hline \\
\end{tabular*}
\end{center}
\end{table*}

\begin{table*}[t]
\begin{center}
Table~9.\hspace{4pt} Model parameters for HR10\\
\vspace{6pt}
\begin{tabular*}{300pt}{@{\hspace{\tabcolsep}
\extracolsep{\fill}}p{3pc}ccccccc}
\hline\hline\\ [-10pt]
\multicolumn{1}{c}{Name} &
\multicolumn{1}{c}{$t$ [Myr]} &
\multicolumn{1}{c}{$\tau_V$} &
\multicolumn{1}{c}{$R$} &
\multicolumn{1}{c}{$\eta$} &
\multicolumn{1}{c}{Extinction curve} &
\multicolumn{1}{c}{Reduced $\chi^2$} \\
\hline \\[-20pt]
model 1  & 300  & 7.0 & 1.0 &1000  & MW & 1.67\\
model 2  & 100  & 5.0 & 3.0 &1000  & MW & 3.59\\
model 3  & 1000  & 20. & 0.1& 1000 & MW & 2.08\\
\hline \\
\end{tabular*}
\end{center}
\end{table*}

\begin{table*}[t]
\begin{center}
Table~10.\hspace{4pt} Model parameters for NGC2768\\
\vspace{6pt}
\begin{tabular*}{300pt}{@{\hspace{\tabcolsep}
\extracolsep{\fill}}p{3pc}ccccccc}
\hline\hline\\ [-10pt]
\multicolumn{1}{c}{Name} &
\multicolumn{1}{c}{$t$ [Gyr]} &
\multicolumn{1}{c}{$\tau_V$} &
\multicolumn{1}{c}{$R$} &
\multicolumn{1}{c}{$\eta$} &
\multicolumn{1}{c}{Extinction curve} &
\multicolumn{1}{c}{Reduced $\chi^2$} \\
\hline \\[-20pt]
model 1  &  12$^a$ & 0.3 & 12 & 1 & MW$^a$ & 1.64\\
model 2  & 12$^a$  & 0.035 & 0.7 & 1000 & MW$^a$ & 1.69\\
\hline \\
\end{tabular*}
\end{center}
\vspace{-16pt}
\par\noindent
$^a$ assumed parameters\\
\end{table*}

\newpage

\begin{figure}[b]
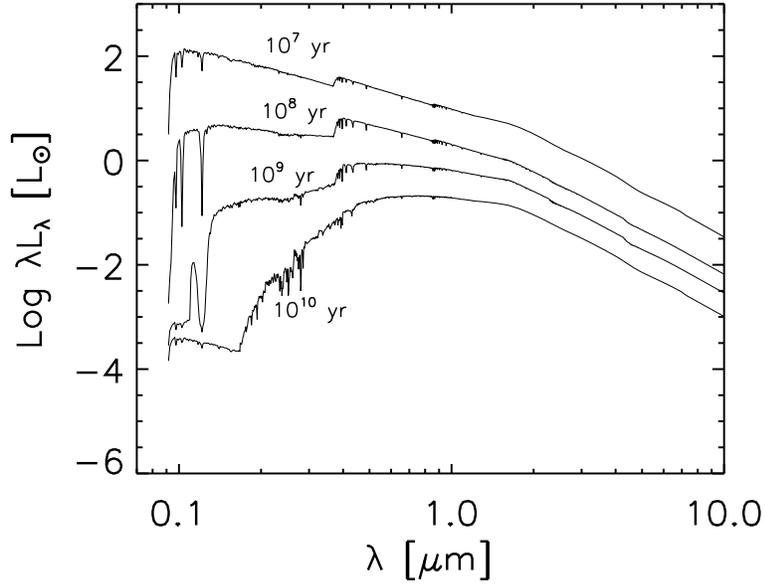

  \begin{center}
  \FigureFile(10cm,8cm){fig1.epsi}
  \end{center}
  \caption{
The spectral energy distributions (SEDs) of the simple stellar
populations (SSP) (Kodama \& Arimoto 1997) are given for ages: 10 Myr
(top), 100 Myr, 1 Gyr, and 10 Gyr (bottom). 
The luminosity is given for the SSP with the initial mass (i.e. 
mass at zero age) of 1 M$_\odot$. 
}
\end{figure}

\begin{figure}
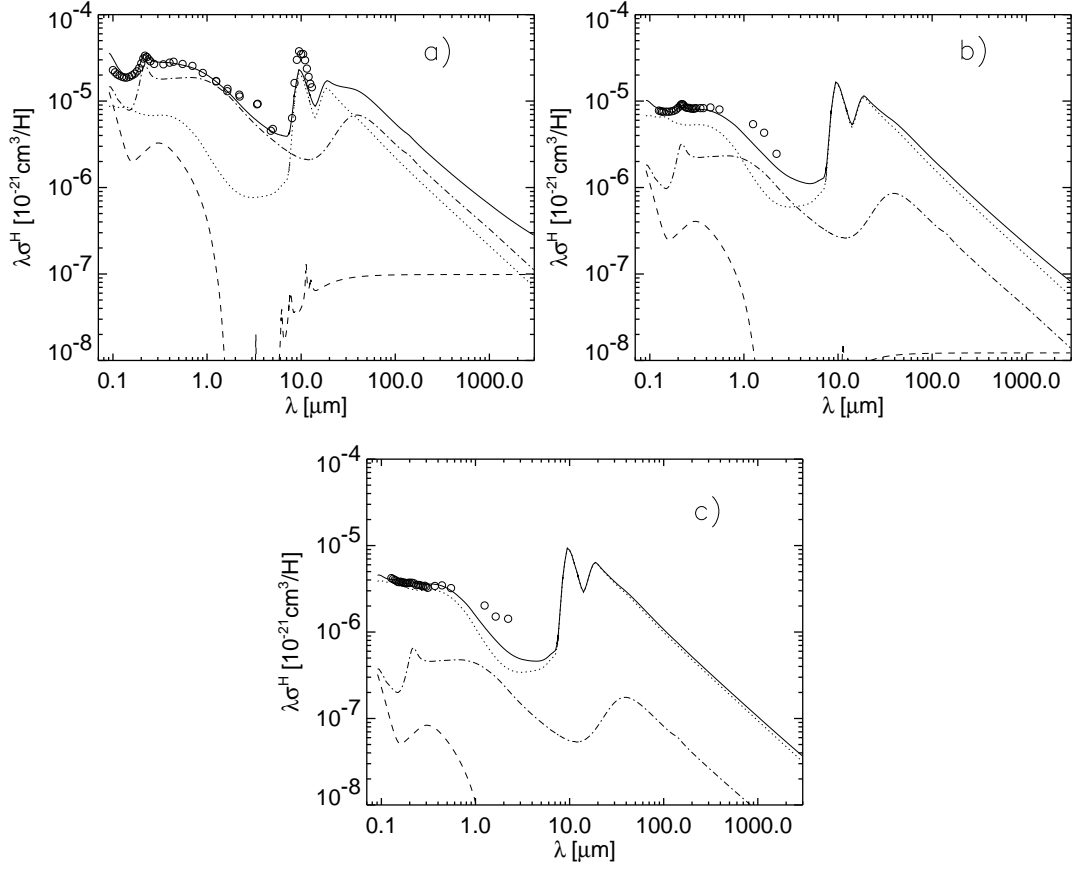

  \begin{center}
  \FigureFile(7cm,6cm){fig2a.epsi}
  \FigureFile(7cm,6cm){fig2b.epsi}
  \FigureFile(7cm,6cm){fig2c.epsi}
  \end{center}
  \caption{
The MW, LMC and SMC type extinction curves per hydrogen atom are shown in
panel a), b) and c), respectively.
The dot-dashed curve is for graphite, dotted for silicate,
dashed for PAH, and solid for total extinction. The average
extinction curves of each type are taken from Pei (1992) and
Rieke \& Lebofsky (1985), and are depicted by open circles.
}
\end{figure}

\begin{figure}
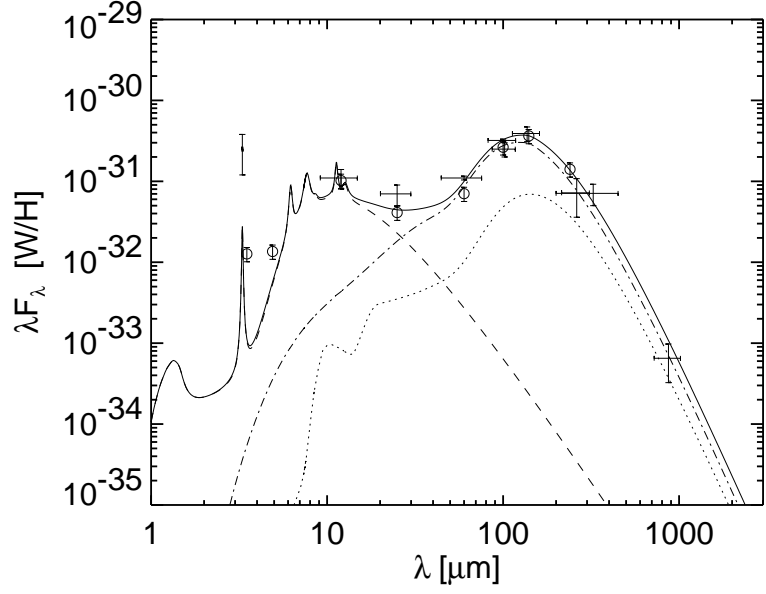

  \begin{center}
  \FigureFile(10cm,0cm){fig3.epsi}
  \end{center}
  \caption{
The spectra of cirrus emission in the MW. The line symbols are
the same as in Figure 2. COBE observations are represented by open
circles (Dwek et al. 1997). The other data are taken from
D\'esert et al. (1990) and references therein. 
}
\end{figure}

\begin{figure}
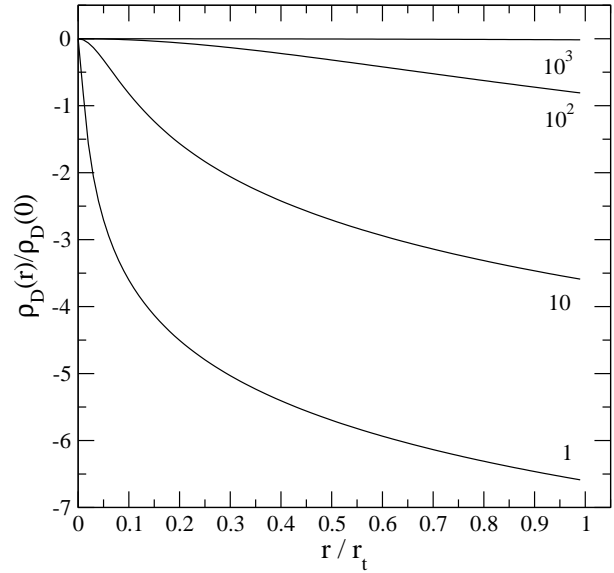

  \begin{center}
  \FigureFile(8cm,8cm){fig4.epsi}
  \end{center}
  \caption{
The normalized dust density profiles given for various $\eta$ values. 
The horizontal axis is normalized by $r_t$ (=$r_{t,S}$). 
}
\end{figure}

\begin{figure}
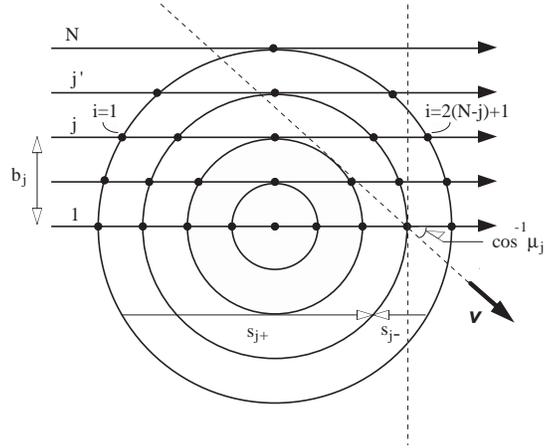

  \begin{center}
  \FigureFile(8cm,8cm){fig5.epsi}
  \end{center}
  \caption{
Scheme of dusty galaxy. The intensity is calculated
and stored at the positions of filled circles. $b_j$ indicates
the impact parameter for the $j$-th ray. The angular integration
at $r_{j'}$, where $j'=j+1$, is performed by using the rays
depicted by dashed lines and that with $j=1$.
The cosine grid $\mu_j$ at $r_{j'}$ is defined by
\mbox{\boldmath $v$} and a radius vector.
}
\end{figure}

\begin{figure}
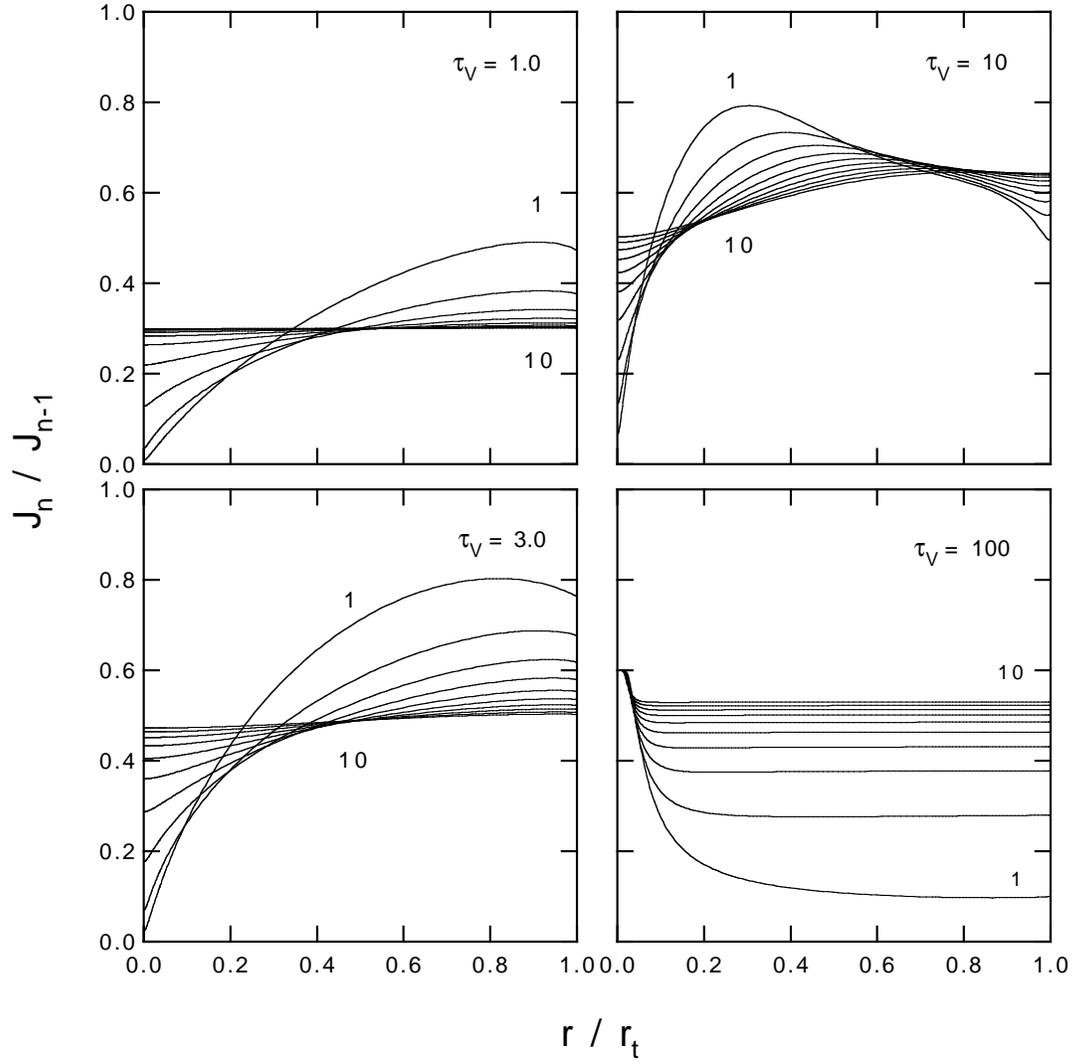

  \begin{center}
  \FigureFile(14cm,12cm){fig6.epsi}
  \end{center}
  \caption{
A mean intensity ratio of scattering terms $J_n/J_{n-1}$,
where $J_n$ and $J_{n-1}$ are $n$ and $n-1$ times scattered
light, respectively. The adopted parameters are $\eta = 1000$,
$\tau_V = 1.0, 3.0, 10, 100$, and $n=1$ to 10 with
$\Delta n =1$.
}
\end{figure}

\begin{figure}
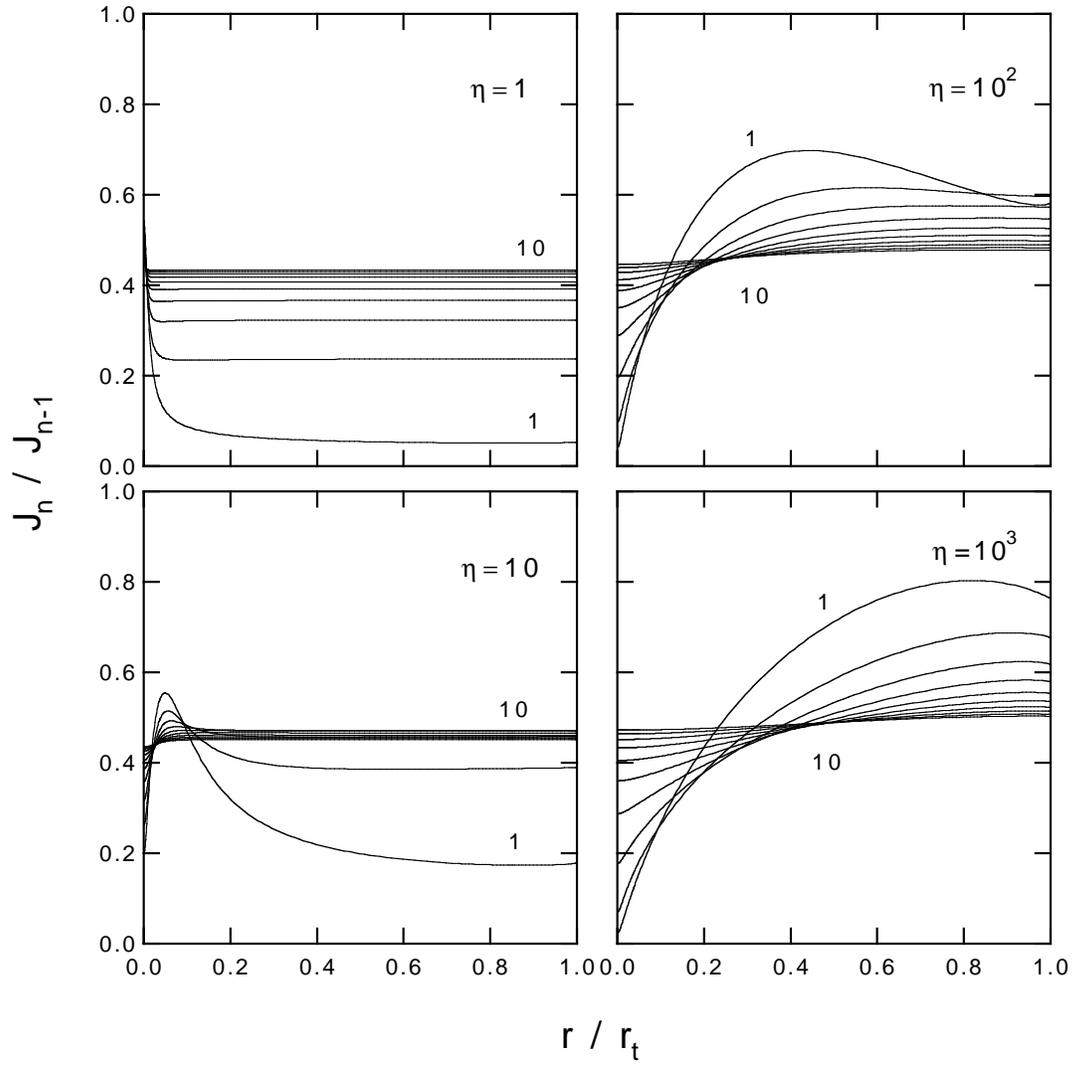

  \begin{center}
  \FigureFile(14cm,12cm){fig7.epsi}
  \end{center}
  \caption{
The same as in Figure 6, but for $\tau_V = 3.0$.
The adopted parameter $\eta = 1, 10, 100, 1000$ is indicated
at the upper-right corner of each panel.
}
\end{figure}

\begin{figure}
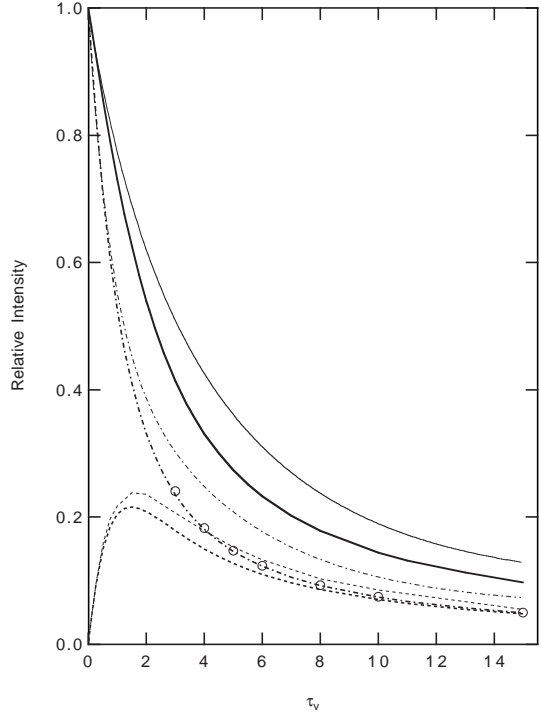

  \begin{center}
  \FigureFile(7cm,6cm){fig8.epsi}
  \end{center}
  \caption{
The fraction of light which escapes from a dust-rich galaxy.
Stars and dust are assumed to distribute homogeneously in this case. 
We adopt $\omega=0.6$. 
The thick lines indicate the results of present model, while
the thin lines are taken from Witt et al. (1992). 
Solid, dot-dashed, and dotted lines show the total light,
the direct stellar light, and the scattered light, respectively.
The analytical solution for the direct stellar light is
shown by open circles.
}
\end{figure}

\begin{figure}
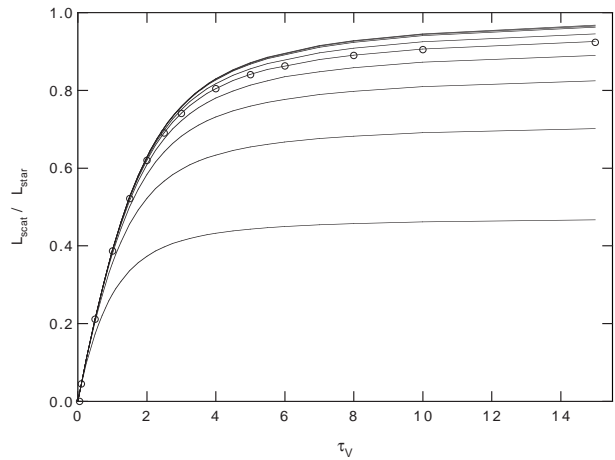

  \begin{center}
  \FigureFile(8cm,7cm){fig9.epsi}
  \end{center}
  \caption{
The ratio of scattered light to the direct stellar light as a 
function of $\tau_V$
for different number of the scattering terms $n=1$ (bottom) to
$n=10$ (top) with $\Delta n=1$. Note that lines for $n \ge 7$ are
nearly identical. 
The results of Witt et al. (1992) are plotted as open circles.
}
\end{figure}

\begin{figure}
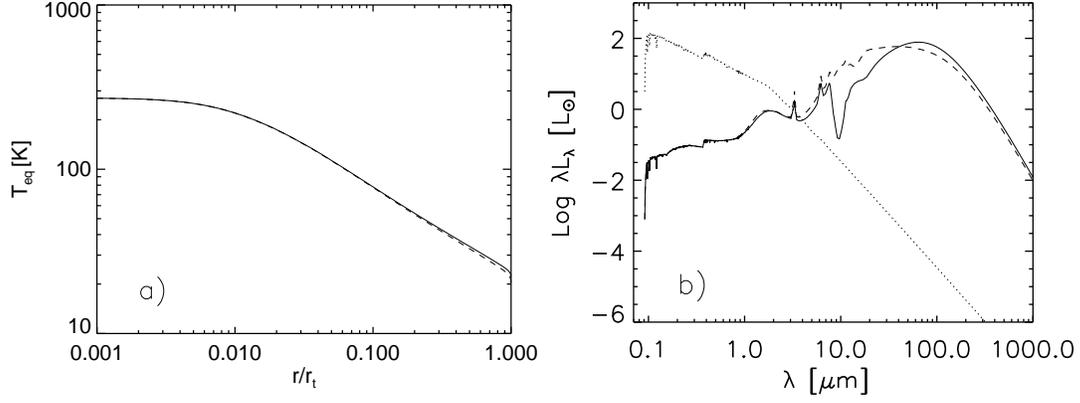

  \begin{center}
  \FigureFile(7cm,6cm){fig10a.epsi}
  \FigureFile(7cm,6cm){fig10b.epsi}
  \end{center}
  \caption{
The effects of self-absorption.
(a) The temperature distributions of graphite BG with $a=0.15~ \mu$m.
The solid and dashed lines indicate $T_{eq}(r)$ with and without
the self-absorption, respectively.
(b) The SEDs. The solid and dashed lines indicate $\lambda L_{\lambda}$
with and without the self-absorption, respectively.
The dotted line gives the intrinsic SED. The adopted parameters are
$\eta = 1000$, $\tau_V = 30.0$, $t = 10$ Myr, and
the SMC extinction curve.
}
\end{figure}

\begin{figure}
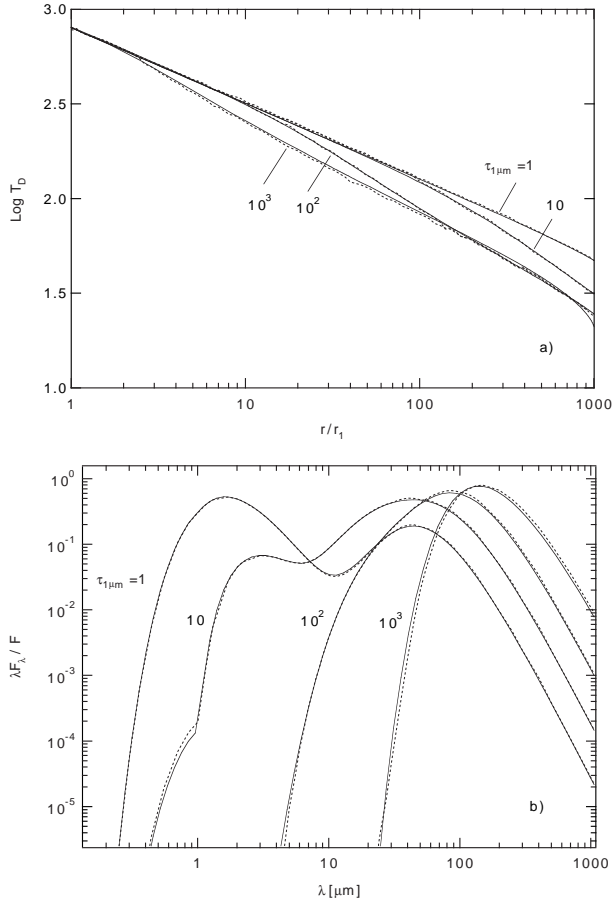

  \begin{center}
  \FigureFile(8cm,18cm){fig11.epsi}
  \end{center}
  \caption{
A comparison with output of DUSTY code by Ivezi\' c et al. (1997).
Solid and dotted lines indicate the results of present model and those
of Ivezi\' c et al., respectively.
(a) The temperature distribution of dust. The radius is given in units
of a radius of central dust free cavity, $r_1$.
(b) The SEDs normalized by an integrated flux, $F$.
}
\end{figure}
\begin{figure}
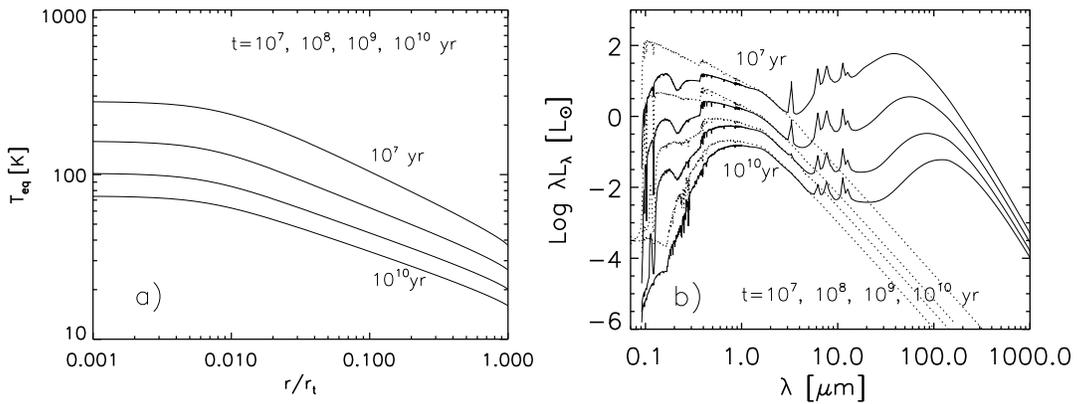


  \begin{center}
  \FigureFile(7cm,6cm){fig12a.epsi}
  \FigureFile(7cm,6cm){fig12b.epsi}
  \end{center}
  \caption{
The effects of age. Parameters, except for the age, are the
same as for the standard model (see text).
(a) The temperature distribution of graphite BG with $a=0.15~ \mu$m
for 10 Myr, 100 Myr, 1 Gyr, and 10 Gyr old.
(b) The same as (a), but for the SEDs. Dotted lines give the intrinsic SEDs.
}
\end{figure}

\begin{figure}
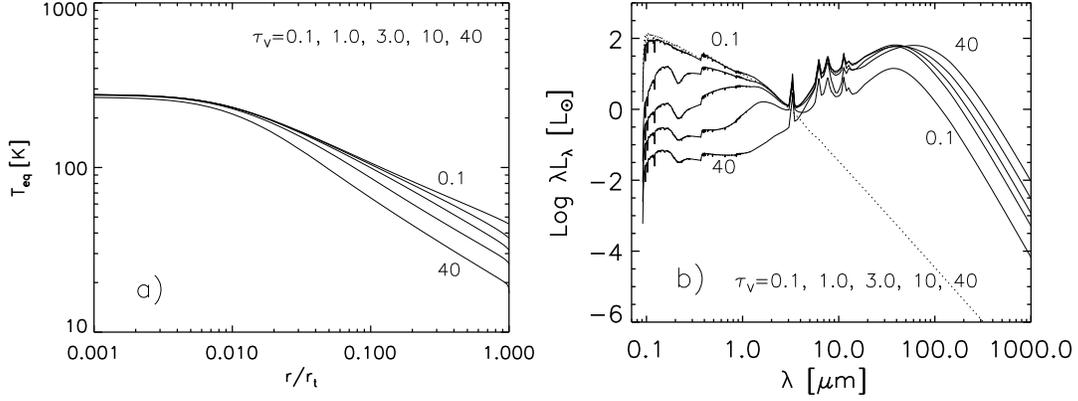

  \begin{center}
  \FigureFile(7cm,6cm){fig13a.epsi}
  \FigureFile(7cm,6cm){fig13b.epsi}
  \end{center}
  \caption{
The effects of optical depth. Parameters, except for $\tau_V$, are the
same as for the standard model (see text).
(a) The temperature distribution of graphite BG with $a=0.15~ \mu$m for
$\tau_V= 0.1$ (top), 1.0, 3.0, 10 and 40 (bottom).
(b) The same as (a), but for the SEDs. The dotted line shows the 
intrinsic SED.
}
\end{figure}
\begin{figure}
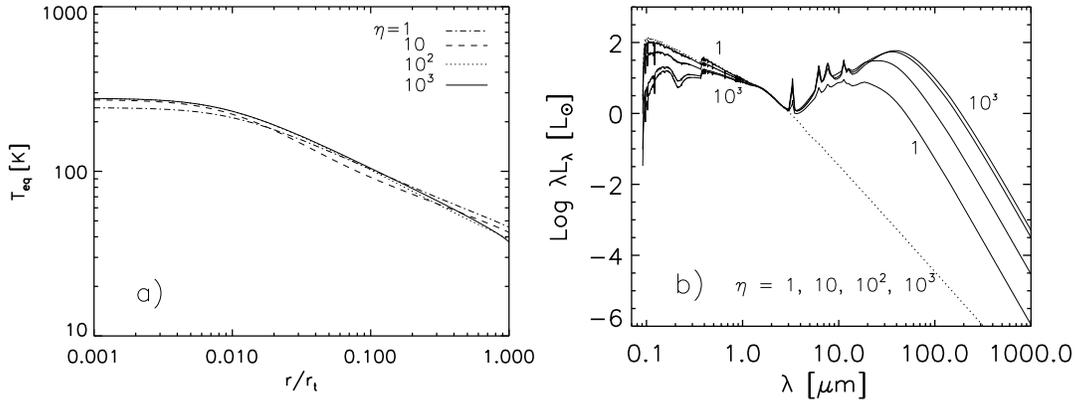


  \begin{center}
  \FigureFile(7cm,6cm){fig14a.epsi}
  \FigureFile(7cm,6cm){fig14b.epsi}
  \end{center}
  \caption{
The effects of dust geometry. Parameters, except for $\eta$, are the
same as for the standard model (see text). (a) The temperature distribution
of graphite BG with $a=0.15~ \mu$m for $\eta=1$ (dot-dashed line),
10 (dashed line), $100$ (dotted line), and $1000$ (solid line).
(b) The same as (a), but for the SEDs. The dotted line shows the intrinsic SED.
Figures indicate $\eta$ values which increase from top to bottom for the
stellar emission, but from bottom to top in FIR-submm range.
}
\end{figure}
\begin{figure}
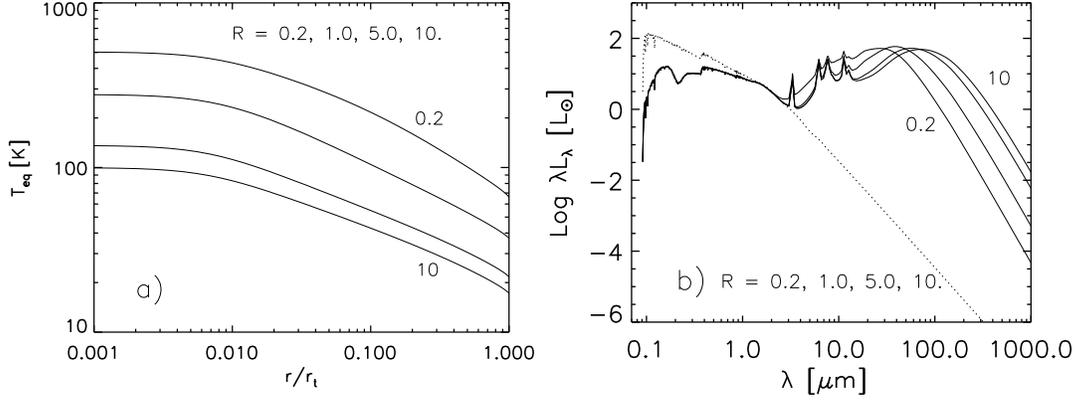


  \begin{center}
  \FigureFile(7cm,6cm){fig15a.epsi}
  \FigureFile(7cm,6cm){fig15b.epsi}
  \end{center}
  \caption{
The effects of system size. Parameters, except for $R$, are the
same as for the standard model (see text).
(a) The temperature distribution of graphite BG with $a=0.15~ \mu$m
for $R=0.2$ (top), 1.0, 5.0 and 10.0
(bottom). (b) The same as (a), but for the SEDs. The dotted line
indicates the intrinsic SED.
}
\end{figure}

\begin{figure}
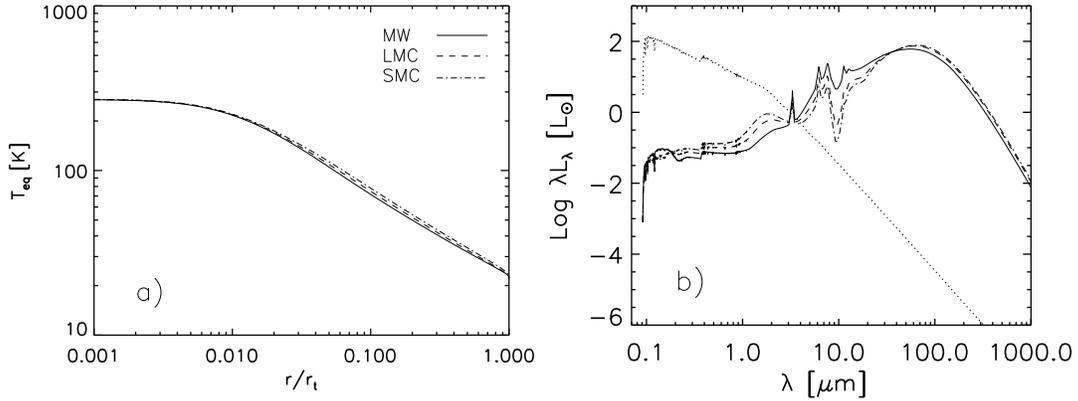

  \begin{center}
  \FigureFile(7cm,6cm){fig16a.epsi}
  \FigureFile(7cm,6cm){fig16b.epsi}
  \end{center}
  \caption{
The effects of extinction curve. Parameters, except for $\tau_V =30$
and the extinction curve, are the same as for the standard model (see text).
(a) The temperature distribution of graphite BG with $a=0.15~ \mu$m
for the MW (solid line), the LMC (dashed line), and the SMC
(dot-dashed line) extinction curves. (b) The same as (a), but for
the SEDs. The dotted line indicates the intrinsic SED.
}
\end{figure}

\begin{figure}
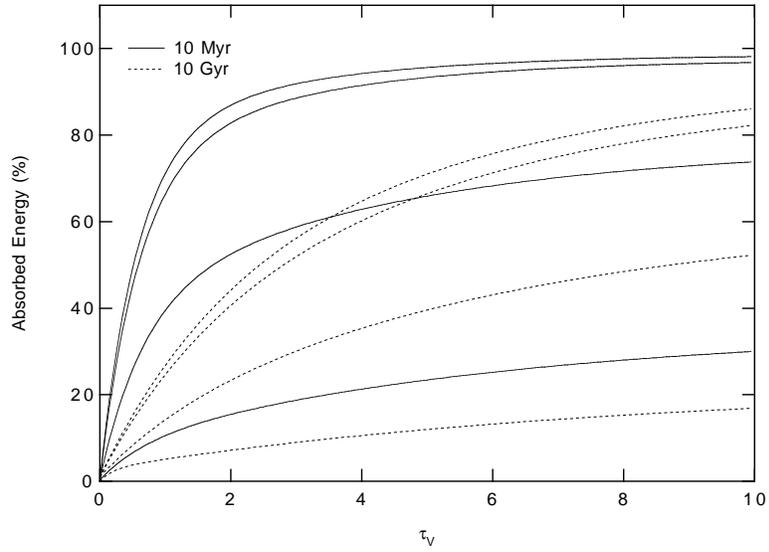

  \begin{center}
  \FigureFile(10cm,8cm){fig17.epsi}
  \end{center}
  \caption{
The absorbed energy fraction as a function of optical depth.
Solid and dotted lines correspond to 10 Myr and 10 Gyr,
respectively. The four lines for each age correspond to
geometry with $\eta = 1$ (bottom), 10, $100$, and $1000$ (top).
}
\end{figure}

\begin{figure}
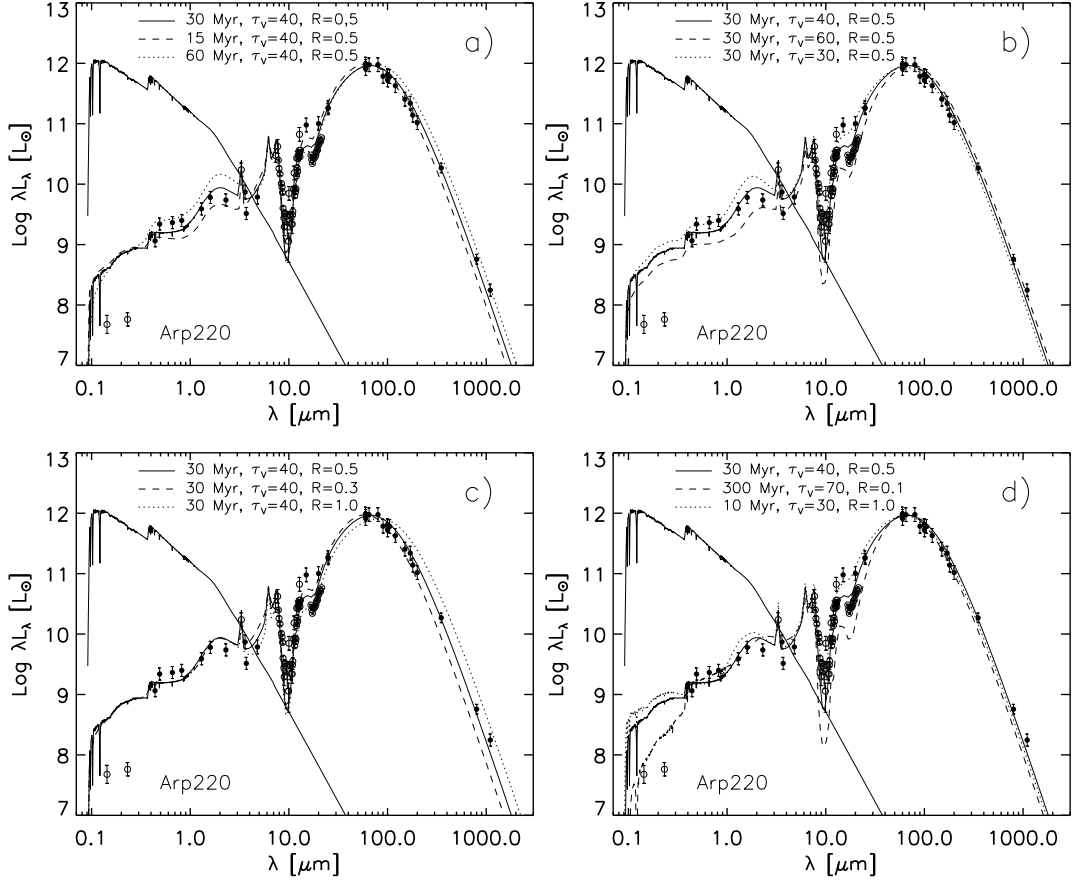

  \begin{center}
  \FigureFile(7cm,6cm){fig18a.epsi}
  \FigureFile(7cm,6cm){fig18b.epsi}
  \FigureFile(7cm,6cm){fig18c.epsi}
  \FigureFile(7cm,6cm){fig18d.epsi}
  \end{center}
  \caption{
The results of SED fitting for Arp220.
The SMC type extinction curve and $\eta = 1000$ are adopted for all models.
The other parameters for each model are denoted in the figure.
We use solid circles for data which are used to calculate the value 
of $\chi^2$,
otherwise open circles are adopted. The upper solid line at UV-NIR region
is the intrinsic stellar SED with the age of 30 Myr. In all panels, the
best-fitting model is indicated by solid lines.
Horizontal axis is the rest frame wavelength.
(a) The SED models with the same parameters, but age,
as that of best-fitting model indicated by solid line.
(b) The same as (a), but $\tau_V$ is varied, instead of age.
(c) The same as (a), but $R$ is varied, instead of age.
(d) The fitting model with the fixed age (10 and 300 Myr), shown
with the best-fitting model.
}
\end{figure}

\begin{figure}
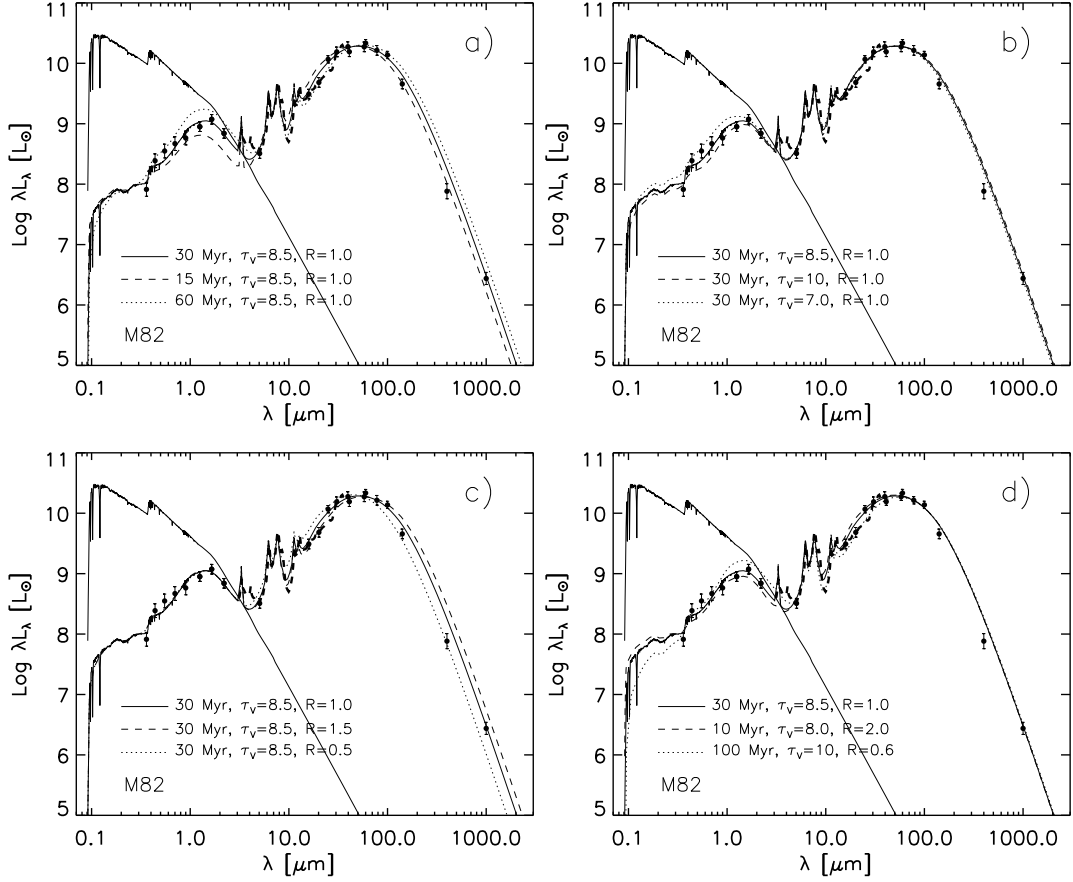

  \begin{center}
  \FigureFile(7cm,6cm){fig19a.epsi}
  \FigureFile(7cm,6cm){fig19b.epsi}
  \FigureFile(7cm,6cm){fig19c.epsi}
  \FigureFile(7cm,6cm){fig19d.epsi}
  \end{center}
  \caption{
The same as Figure 19, but for M82.
LMC type extinction curve and $\eta = 1000$ are adopted for all models.
Thick dashed line is for the ISO-SWS spectrum.
The upper solid line at UV-NIR region
is the intrinsic stellar SED with the age of 30 Myr.
}
\end{figure}

\newpage

\begin{figure}
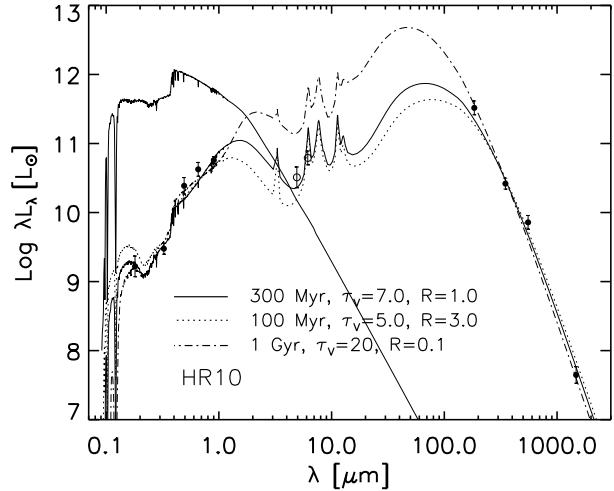

  \begin{center}
  \FigureFile(8cm,7cm){fig20.epsi}
  \end{center}
  \caption{
The same as Figure 18d, but for the high redshift galaxy, HR10.
ISO data are indicated by open circles.
The MW extinction curve and $\eta = 1000$ are adopted for all models.
The upper solid line in UV-NIR is
the intrinsic stellar SED with $t = 300$~Myr.
}
\end{figure}

\begin{figure}
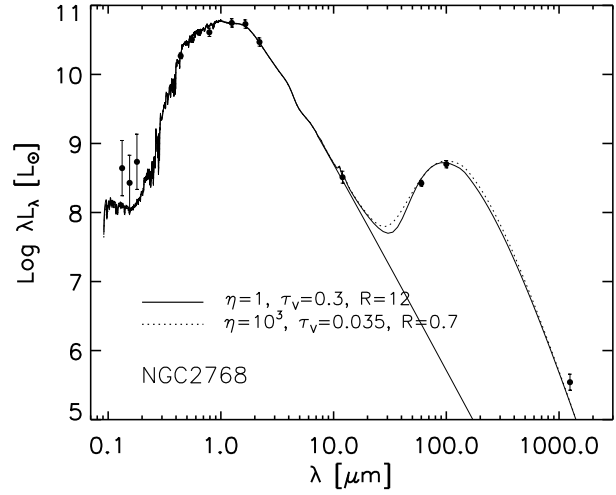

  \begin{center}
  \FigureFile(8cm,7cm){fig21.epsi}
  \end{center}
  \caption{
The results of SED fitting for the elliptical galaxy, NGC2768.
The MW type extinction curve and $t = 12$~Gyr are adopted for all models.
The stellar emission (12 Gyr) in FIR is indicated by lower solid line.
}
\end{figure}

\newpage
\bigskip

\newpage

\end{document}